\documentclass[a4paper,twocolumn,showkeys]{revtex4}  
\usepackage{latexsym}
\usepackage[dvipdfm,colorlinks=true,linkcolor=blue,anchorcolor=blue,citecolor=blue]{hyperref}
\usepackage[hmargin={1cm,1cm}, vmargin={2cm,2cm}]{geometry}
\usepackage{float}
\usepackage{graphicx,amsfonts,amsmath,amssymb,amsthm,amscd}
\usepackage{epsfig}
\usepackage{bm}
\usepackage{xr}
\usepackage{indentfirst}
\usepackage{appendix}
\usepackage{changepage}
\usepackage{multirow}

\newcommand{\tabincell}[2]{\begin{tabular}{@{}#1@{}}#2\end{tabular}}

\date{\today}

\begin{document}

\title{\bf Biochemical pathways of forward/backward steps of motor protein kinesin-1}

\author{Beibei Shen and Yunxin Zhang} \email[Email: ]{xyz@fudan.edu.cn}
\affiliation{Shanghai Key Laboratory for Contemporary Applied Mathematics, School of Mathematical Sciences, Fudan University, Shanghai 200433, China.}

\begin{abstract}
Kinesin-1 is an ATP-driven, two-headed motor protein that transports intracellular cargoes along microtubule.
Based on recent experimental observations, we formulate a mechanochemical model for it, in which  forward/backward/futile cycle of kinesin-1 can be realized through multiple biochemical pathways. Our results show that both forward and futile cycles consist of two  biochemical pathways, while backward cycle may be realized through six possible pathways. Backward motion of kinesin-1 is mainly through backward sliding along microtubule when it is in semi-detach state, and usually coupled with ATP hydrolysis. Under low external load, about 80\% of ATP is wasted by kinesin-1, and this proportion of waste is almost independent of ATP concentration. At high ATP concentration or under high external load, both heads of kinesin are always in ATP or ADP$\cdot$Pi binding state and bind to microtubule tightly. But at low ATP concentration and low load, kinesin mainly stays at the one head bound state. In contrast to mean run length, the mean run time of kinesin-1 along microtubule decreases with ATP concentration. Unless external load is near the stall force, the motion of kinesin-1 is almost deterministic, not as predicted by usual Browian ratchet models.
\end{abstract}



\maketitle

\section{Introduction}
Kinesin-1 plays a central role in the intracellular transport of various vesicles and organelles \cite{Howard2001,Vale2003,Schliwa2003,Kolomeisky2015}. Each kinesin dimer consists of two motor domains (heads), which are connected to a coiled-coil stalk through a $\sim$14-amino-acid-long sequence known as the neck linker (NL) \cite{Case2000,Jawdat2003,Tomishige2006,Kalchishkova2008,Hariharan2009Insights}.
Kinesin proceeds unidirectionally toward the plus ends of microtubule (MT) by an asymmetric hand-over-hand fashion, hydrolyzing an ATP molecule for each 8 nm step \cite{Asbury2003,sbb1,Schnitzer1997,sbb2,sbb3,Higuchi2004,Guydosh2009}. Motion of kinesin is significantly processive \cite{sbb7,Toprak2009}, taking about 100 steps before detaching from the MT \cite{Howard2001,clancy2011universal,Takayuki2018}.
At low loads, kinesin-1 always steps forward to the plus ends of MT, while at high loads, it can also result in backward steps \cite{svoboda1994force,Visscher1999,Block2003}.

In previous studies, kinds of theoretical models have been presented to understand the mechanism of kinesin motion; see \cite{Parrondo2002,Kolomeisky2007,Chowdhury20131,Mugnai2020} for details. Generally, the motion of kinesin can be regarded as a Markov process with multiple biophysical and biochemical states, and in steady-state, meaningful quantities can be obtained by using Fokker-Planck equation, or master equation, or Langevin dynamics, or any other complicated hybrid models \cite{Kampen2007,Gardiner2010}. Among most of these models, a backward step of kinesin is usually thought to occur by directional reversal of a forward step and therefore happens with ATP synthesis.

Generally, both forward and backward steps of kinesin may be accomplished through different biochemical pathways, and backward motion of kinesin may be resulted simply from pure biophysical slips along MT \cite{sbb4}. According to a mechanochemical model, our study shows that there are two biochemical pathways for forward step. At low ATP concentration and low external load, kinesin will spend more time waiting for ATP, and its MT-bound head might be in the nucleotide-free state.
However, at high ATP concentration or high external load, all MT-bound heads are usually in ATP or ADP$\cdot$Pi binding state, and ATP binding to the front head of kinesin is earlier than the release of Pi and the detachment of the trailing head from MT.
There are altogether six possible pathways for backward motion, unless at some extreme conditions the backward motion of kinesin is through backward slips along MT when it is in semi-detach state, and usually coupled with ATP hydrolysis. Directional reversal of forwarding pathways is only non-negligible when the ATP molecule is  scarce while the external load is very high. One surprising finding is that about 80\% of ATP is wasted by kinesin, and this proportion of waste is almost independent of ATP concentration.
\begin{figure}[htbp]
\includegraphics[scale=0.28]{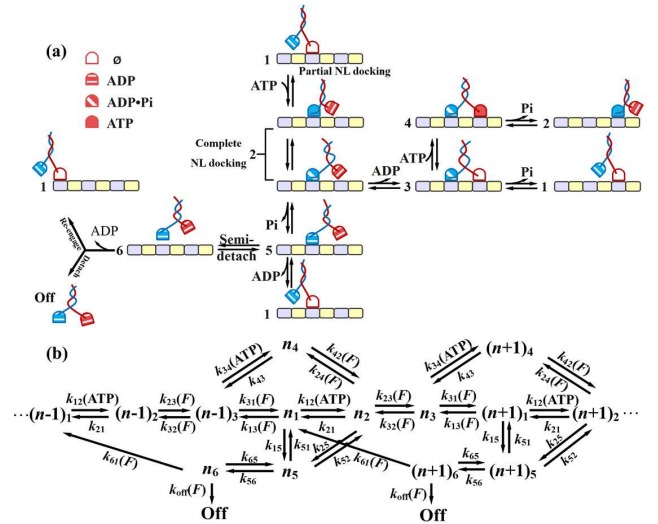}\\
\caption{\label{fig:1} The mechanochemical model of kinesin-1. \textbf{(a)} The depiction of detailed configurations of the two heads of kinesin-1. There are six possible configurations at each binding site of MT.
\textbf{(b)} Mechanochemical model used in this study to describe the periodic motion of kinesin-1, where $n_j$ denotes the state that kinesin is at position $n$ and in configuration $j$, with $j=1,2,\cdots,6$.}
\end{figure}

\section{The Mechanochemical model of kinesin-1}
Based on experimental observations in \cite{sbb4}, this study uses the mechanochemical model as depicted in Fig.~\ref{fig:1}. The biochemical cycle of kinesin is assumed to begin with the one-head-bound, ATP-waiting state, with the nucleotide-free front head is tightly bound to the MT while the ADP-bound rear head unbound \cite{sbb5}. ATP binding induces NL docking and ATP hydrolysis completes NL docking \cite{Case2000,Tomishige2006,Kalchishkova2008} (see Fig.~\ref{fig:1}{\bf(a)} 1$\to$2). In either ATP or ADP$\cdot$Pi bound states, the head binds tightly to the MT, so these two states are assumed to be the same and denoted as state 2 in Fig.~\ref{fig:1}. Due to NL docking, the unbound head of kinesin swings forward and becomes the front head. ADP release facilitates front head binding to MT, leading kinesin changes from state 2 to state 3, generating a two-heads-bound state. At low ATP concentrations, the release of phosphate ion Pi from the rear head usually occurs before an ATP molecule reaches the newly bound front head. Then dissociation of the ADP-bound rear head from the MT leads kinesin to return to biochemical state 1 again. However, if the ATP concentration reaches saturating, after ADP releases from the front head, another ATP molecule might bind to the nucleotide-free head immediately, leading kinesin to enter the biochemical state 4. Next, the release of Pi from the rear head will lead kinesin to return to state 2. In either case, 1$\to$2$\to$3$\to$1 or 2$\to$3$\to$4$\to$2, one biochemical cycle of kinesin is completed once the ADP bound rear head dissociates from the MT. With completion of one biochemical cycle, kineisn will move forward about 8 nm along the MT, and is ready to start a new mechanochemical cycle again \cite{Schnitzer1997,sbb2,Higuchi2004}.

Recent experiments discovered that the dwell time for forwarding step is shorter than that for backward step, and dwell times for backward step and detachment are almost the same \cite{sbb4}. Here dwell time is the waiting time preceding each step, consisting of the time spent waiting for ATP to bind and the time is taken to process ATP and complete the coupled mechanical step.
This implies that backward step originates from a different and later state in the biochemical cycle than the one that generates forward steps, and backward step and detachment might emanate from the same state. Therefore, our model assumes that forward step can only occur before Pi release from the ADP$\cdot$Pi bound head (see state 2 in Fig.~\ref{fig:1}), whereas backward step/slip and detachment can only occur after Pi release.
The release of Pi leads to a weakly bound ADP state (state 5). With increasing external load, forward step of motor becomes difficult to complete before Pi release, while backward step may occur from state 5. If the rear head releases its ADP before it is semi-detached from the MT, kinesin in state 5 can return to state 1. Evidently, one ATP is hydrolyzed during cycle 1$\to$2$\to$5$\to$1 but without output of mechanical work, which we called {\it futile} cycle in this study. As concluded in \cite{sbb8}, the inclusion of {\it futile} cycles is experimentally meaningful and theoretically essential to make the model fit the experimental data well.

Semi-detach state 6 is the weakest bounding state in the biochemical cycle of kinesin, and any backward slipping necessarily requires passage through this state. From state 6, kinesin may dissociate completely from MT, and enter into the detachment state, denoted as Off in Fig.~\ref{fig:1}. More importantly, as described in \cite{sbb4}, kinesin in state 6 may also slip backward along the MT. For simplicity, our model assumes that, unless it is completely detached from the MT, kinesin always slips backward $d=8$ nm along the MT in each slipping process, then releases the ADP from its semi-detached head and returns to state 1. This assumption can be relaxed to allow kinesin to slip backward any distance of integer multiple of 8 nm. But results of fitting to experimental data show that the load free rate $k_{61}^0$ is very small, so the possibility of backward sliding is small, and this simplification is reasonable, see Tab.~\ref{tab:1}. Meanwhile, kinesin in semi-detach state 6 may also slide forward along the MT. But, since the probability of kinesin in state 6 is very small (see Fig.~\ref{fig:3}), compared with the total forward probability flux, the flux produced by forwarding sliding is negligible.

The model used in this study is similar to the one constructed in \cite{sbb4}. One of the key features distinguishing it from earlier ones is the inclusion of semi-detach state 6, from which the motor can slip backward to state 1 or dissociate from MT completely. Generally, biochemical cycles of kinesin in this model can be classified into three categories, {\bf (i)} {\it forward} cycle coupled with one forward mechanical step, {\bf (ii)} {\it backward} cycle coupled with one backward mechanical step, and {\bf (iii)} {\it futile} cycle with one ATP hydrolyzed but without change of position.

As illustrated in Fig.~\ref{fig:1}\textbf{(a)}, biochemical transitions 2$\rightleftharpoons$3, 3$\rightleftharpoons$1, and  4$\rightleftharpoons$2 are coupled with head attachment/detachment to/from MT binding site, which may result in the change of mass center of kinesin. So rates of these transitions, {\it i.e.,} $k_{23}$, $k_{32}$, $k_{31}$, $k_{13}$, $k_{42}$, $k_{24}$, are assumed to be external load $F$ dependent. Similar as in \cite{fisher2001simple,Zhang2012}, this study assumes that
\begin{equation}
\begin{aligned}
&k_{23}(F)=k_{23}^{0}e^{\frac{-\delta_{23}Fd}{k_{B}T}},\quad
k_{32}(F)=k_{32}^{0}e^{\frac{\delta_{32}Fd}{k_{B}T}},\\
&k_{31}(F)=k_{31}^{0}e^{\frac{-\delta_{31}Fd}{k_{B}T}},\quad
k_{13}(F)=k_{13}^{0}e^{\frac{\delta_{13}Fd}{k_{B}T}},\\
&k_{42}(F)=k_{42}^{0}e^{\frac{-\delta_{42}Fd}{k_{B}T}},\quad
k_{24}(F)=k_{24}^{0}e^{\frac{\delta_{24}Fd}{k_{B}T}},
\end{aligned}
\end{equation}
where $k_{ij}^{0}$ are load free transition rates, $F$ is the external load (positive if it points to the minus end of MT), $k_B$ is the Boltzmann constant, $T$ is the absolute temperature, $d=8$ nm is the step size of kinesin. $\delta_{ij}\geq0$ is called {\it load distribution factor} that reflects how the external load affects the rate of transition from state $i$ to state $j$, which satisfies
$$
\delta_{23}+\delta_{32}+\delta_{31}+\delta_{13}=1, \quad \delta_{23}+\delta_{32}+\delta_{42}+\delta_{24}=1.
$$

Since transitions 4$\to$2 and 3$\to$1 describe the same biochemical process, during which phosphate Pi is released and then the trailing head is detached from MT, we let $k_{42}^0=k_{31}^0$ and $k_{24}^0=k_{13}^0$. However, the position of energy barrier between states 4 and 2 might be different from that between states 3 and 1, so load distribution factors $\delta_{42}, \delta_{24}$ are generally different from $\delta_{31}, \delta_{13}$ \cite{fisher2001simple,Howard2001}.

The detachment rate $k_{\rm{off}}$ of kinesin from MT is assumed to be load $F$ dependent and formulated as
$$
k_{\rm{off}}(F)=k_{\rm{off}}^{0}e^{\frac{Fd_{\rm{off}}}{k_{B}T}},
$$
with $d_{\rm{off}}\geq0$ a {\it characteristic distance}. This expression is actually equivalent to the one used previously in \cite{Muller4609,ZhangFisher2010}, where the detachment rate is assumed the be the form $\epsilon_0\exp(|F|/F_d)$ with parameter $F_d$ called detachment force.

It is biophysically reasonable to assume the sliding process from states 6 to 1 is load-dependent, and the rate $k_{61}$ is given as
$$
k_{61}(F)=k_{61}^{0}e^{\frac{Fd}{k_{B}T}},
$$
since each sliding process is always coupled with a backward translocation of $d=8$ nm.

In addition, transitions 1$\to$2 and 3$\to$4 describe ATP binding to the nucleotide-free head of kinesin. So rates $k_{12}$ and $k_{34}$ depend on ATP concentration ${\rm [ATP]}$ in environment. For simplicity, we let
$$
k_{12}({\rm ATP})=k_{12}^{0}{\rm [ATP]}=k_{34}^{0}{\rm [ATP]}=k_{34}({\rm ATP}).
$$
The inverse transitions 2$\to$1 and 4$\to$3 describe the disassociation of ATP molecule from kinesin head. So rates $k_{21}$ and $k_{43}$ are independent of load $F$ and ATP concentration [ATP]. Also for simplicity, we let $k_{21}=k_{43}$ in our model.

Rates $k_{25}, k_{52}, k_{51}, k_{15}$ describe the release/recruiting of phosphate Pi or ADP. For simplicity, we always assume the change of concentrations of ADP and Pi in environment are negligible. So, all these rates are assumed to be constants, though more sophisticated methods can be used to handle their dependence on [ATP] \cite{fisher2001simple}.

\section{Theoretical analysis of the model}
\subsection{Mean velocity and diffusion coefficient}
According to the model illustrated in Fig.~\ref{fig:1}\textbf{(b)}, the probability $p_j^n(t)$ of finding kinesin in biochemical state $j$, at binding site $n$, and at time $t$, is governed by the following master equations, {\small
\begin{eqnarray}
	\begin{aligned}
		&\partial_tp_1^n=k_{31}p_3^{n-1}+k_{21}p_2^n+k_{51}p_5^n+k_{61}p_6^{n+1}\\
&\qquad\quad-(k_{12}+k_{13}+k_{15})p_1^n,\\
		&\partial_tp_2^n=k_{12}p_1^n+k_{32}p_3^n+k_{42}p_4^{n-1}+k_{52}p_5^n\\
&\qquad\quad-(k_{21}+k_{23}+k_{24}+k_{25})p_2^n,\\
		&\partial_tp_3^n=k_{23}p_2^n+k_{43}p_4^n+k_{13}p_1^{n+1}-(k_{32}+k_{31}+k_{34})p_3^n,\\
		&\partial_tp_4^n=k_{34}p_3^n+k_{24}p_2^{n+1}-(k_{43}+k_{42})p_4^n,\\
		&\partial_tp_5^n=k_{15}p_1^n+k_{25}p_2^n+k_{65}p_6^n-(k_{51}+k_{52}+k_{56})p_5^n,\\
		&\partial_tp_6^n=k_{56}p_5^n-(k_{61}+k_{65})p_6^n.
	\end{aligned}\label{Eqforward}
\end{eqnarray}}
Note that, the detachment process will be added in the analysis of mean run-length/time. The distance between two adjacent binding sites, {\it i.e.} the step size of kinesin, is $d=8$ nm.

Using Eq.~(\ref{Eqforward}), the Fourier transform of probability $p_{j}^n(t)$, defined by $P_{j}(q, t):= \sum_{n=-\infty }^{\infty}p_{j}^n(t)e^{-iqn}$, satisfies
\begin{equation}
	\partial_t\bm{P}(q, t) =\bm{M}(q) \bm{P}(q,t),
	\label{matrixx}
\end{equation}
where $\bm{P}(q,t)=[P_{1}(q,t),\cdots,P_{6}(q,t)]^{T}$. Matrix $\bm{M}(q)$ depends on all rates $k_{ij}$ labeled in Fig.~\ref{fig:1}\textbf{(b)} and the variable $q$, see Supplemental Material for its detailed elements.

Taking Laplace transform of Eq.~\eqref{matrixx}, we get a $6$th order characteristic polynomial in $s$,
\begin{equation}
	\left | s\bm{I}-\bm{M}(q)  \right |=s^{6}+\cdots +\alpha (q)s^{2}+\beta (q)s+\gamma (q),
	\label{RR}
\end{equation}
with $\bm{I}$ the identity matrix. As stated in \cite{Koza1999,Koza2000,sbb9}, to obtain expressions of the mean velocity $v$, diffusion coefficient $D$, coefficients of the last three terms of the characteristic polynomial are enough. It can be shown that, see Supplemental Material for detailed derivation,
\begin{equation}
	v=-\frac{id\dot{\gamma}(0)}{\beta (0)},
	\label{vvv}
\end{equation}
\begin{equation}
D=\frac{d^2}{2}\left(\frac{\ddot{\gamma}(0)}{\beta(0)}-\frac{2\dot{\gamma}(0)\dot{\beta}(0)}{\beta(0)^{2}}+\frac{2\alpha(0)\dot{\gamma}(0)^{2}}{\beta(0)^{3}}\right),
		\label{vx}
\end{equation}
where \lq$\cdot$' denotes the derivative with respect to $q$. From Eqs.~\eqref{vvv} and \eqref{vx}, the randomness defined as $r = 2D/vd$ is
\begin{equation}\label{Eqr}
	r=i\left ( \frac{\ddot{\gamma}(0)}{\dot{\gamma}(0)}-2\frac{\dot{\beta}(0)}{\beta(0)}+2\frac{\alpha(0)\dot{\gamma}(0)}{\beta(0)^{2}}\right).
\end{equation}
Obviously, $v,D$ and $r$ are all functions of transition rates $k_{ij}$, and depend on external load $F$ and ATP concentration [ATP].

\subsection{Mean run length/time along MT}
Let $\rho_i=\lim_{t\to\infty}\sum_{n=-\infty }^{\infty}p_{i}^n(t)$ be the steady state probability of finding the motor in biochemical state $i$, then from Eq.~\eqref{Eqforward}, we have
\begin{equation}
	M\rho=\bm{0},
	\label{EqMX}
\end{equation}
with matrix $M=\bm{M}(0)$, $\rho=[\rho_1,\cdots,\rho_6]^T$, and $\bm{0}$ the zero vector. Generally, $\rho$ can be obtained from Eq.~(\ref{EqMX}) combined with the normalization condition $\sum_{i=1}^{6}\rho_{i}=1$.

Denoting $F_m^{n_{j}}(t)$ the probability density of kinesin, which is initiated from site $n$ and biochemical state $j$ at time $t=0$ and detaches for the first time from site $m$ of MT at time $t$. It can be shown that $F_m^{n_{j}}(t)$ satisfy the following backward master equations, see \cite{Kampen2007, Gardiner2010, Kolomeisky2015, ZhangKolomeisky2018}, {\small
\begin{equation}
	\begin{aligned}
		&\partial_t F_m^{n_{1}}=k_{13}F_m^{(n-1)_{3}}+k_{12}F_m^{n_{2}}+k_{15}F_m^{n_{5}}\\
&\qquad\qquad-(k_{12}+k_{13}+k_{15})F_m^{n_{1}},\\
		&\partial_t F_m^{n_{2}}=k_{21}F_m^{n_{1}}+k_{23}F_m^{n_{3}}+k_{24}F_m^{(n-1)_{3}}+k_{25}F_m^{n_{5}}\\
&\qquad\qquad-(k_{21}+k_{23}+k_{24}+k_{25})F_m^{n_{2}},\\
		&\partial_t F_m^{n_{3}}=k_{31}F_m^{(n+1)_{1}}+k_{32}F_m^{n_{2}}+k_{34}F_m^{n_{4}}\\
&\qquad\qquad-(k_{31}+k_{32}+k_{34})F_m^{n_{3}},\\
		&\partial_t F_m^{n_{4}}=k_{42}F_m^{(n+1)_{2}}+k_{43}F_m^{n_{3}}-(k_{42}+k_{43})F_m^{n_{4}},\\
		&\partial_t F_m^{n_{5}}=k_{51}F_m^{n_{1}}+k_{52}F_m^{n_{2}}+k_{56}F_m^{n_{6}}\\
&\qquad\qquad-(k_{51}+k_{52}+k_{56})F_m^{n_{5}},\\
		&\partial_t F_m^{n_{6}}=k_{61}F_m^{(n-1)_{1}}+k_{65}F_m^{n_{5}}+k_{\textrm{off}}\delta _{nm}\delta(t)\\
&\qquad\qquad-(k_{61}+k_{65}+k_{\textrm{off}})F_m^{n_{6}},
	\end{aligned}
\label{Eqbackward}
\end{equation}}
where $\delta _{nm}=0$ if $n\ne m$, and $\delta _{nn}=1$.
$\delta(t)$ in the last equation means that if kinesin detaches from MT, the first-passage process is accomplished immediately.

From Eq.~(\ref{Eqbackward}), the \textit{conditional} mean run length of kinesin which starts from biochemical state $i$,
$$l_i^{(1)}:=\sum_{m}^{}m\int_{0}^{+\infty}F_m^{0_{i}}(t)\mathrm{d}t,$$
satisfies $K\bm{l}^{(1)}=\bm{b}^{(1)}$, with $\bm{l}^{(1)}=[l_1^{(1)},\cdots,l_6^{(1)}]^T$, $\bm{b}^{(1)}=[k_{13}, k_{24}, -k_{31}, -k_{42}, 0, k_{61}]^T$, and matrix $K$ is similar to the coefficient matrix of Eq.~(\ref{Eqbackward}), for details see Supplemental Material.
The mean run length of kinesin along MT is then obtained by
\begin{eqnarray}\label{EqRunlength}
\langle l\rangle =\sum_{i=1}^{6} \rho_il_i^{(1)}=\rho\cdot\bm{l}^{(1)}.
\end{eqnarray}

In the same way, the second moment of \textit{conditional} run length
$$ l_i^{(2)}:=\sum_{m}^{}m^2P\int_{0}^{+\infty}F_m^{0_{i}}(t)\mathrm{d}t,$$
satisfies $K\bm{l}^{(2)}=2\bm{b}^{(1)}\circ\bm{l}- \bm{b}^{(2)}$, where \lq$\circ$' is the Hadamard product, $\bm{l}^{(2)}=[l_1^{(2)},\cdots,l_6^{(2)}]^T$, $\bm{l}=[l_3^{(1)},l_4^{(1)},l_1^{(1)},l_2^{(1)},0,l_1^{(1)}]^T$, $\bm{b}^{(2)}=[k_{13}, k_{24}, k_{31}, k_{42}, 0, k_{61}]^T$ is the absolute value of $\bm{b}^{(1)}$. So the variance of run length is
\begin{eqnarray}\label{EqVarRunlength}
{\rm Var}(l)&=&\langle l^{2}\rangle -\langle l \rangle^{2}
=\sum_{i=1}^{6}\rho_{i}l_i^{(2)}-\left(\sum_{i=1}^{6} \rho_il_i\right)^2\cr
&=&\rho\cdot\bm{l}^{(2)}-(\rho\cdot\bm{l}^{(1)})^2.
\end{eqnarray}

The mean run time of kinesin initiated from biochemical state $i$ is
$$t_i^{(1)}:=\int_{0}^{+\infty}t\sum_{m}^{}F_m^{0_{i}}(t)\mathrm{d}t.$$
Using Eq.~({\ref{Eqbackward}}), $\bm{t}^{(1)}:=[t_1^{(1)},\cdots,t_6^{(1)}]^T$ satisfies $K\bm{t}^{(1)}=-\bm{e}$ with $\bm{e}=[1,\cdots,1]^T$. So the mean run time of kinesin along MT is
\begin{eqnarray}\label{EqMeanT}
\langle t \rangle =\sum_{i=1}^{6}\rho_{i}t_i^{(1)}=\rho\cdot\bm{t}^{(1)}.
\end{eqnarray}
Similarly, the variance of run time is
\begin{eqnarray}\label{EqVarT}
{\rm Var}(t)&=&\langle t^{2}\rangle -\langle t\rangle^{2}
=\sum_{i=1}^{6}\rho_{i}t_i^{(2)}-\left(\sum_{i=1}^{6} \rho_it_i\right)^2\cr
&=&\rho\cdot\bm{t}^{(2)}-(\rho\cdot\bm{t}^{(1)})^2,
\end{eqnarray}
where $\bm{t}^{(2)}=[t_1^{(2)},\cdots,t_6^{(2)}]^T$ satisfies $K\bm{t}^{(2)}=-2\bm{t}^{(1)}$, see \cite{Gardiner2010}.

\begin{figure}[htbp]	\includegraphics[scale=0.28]{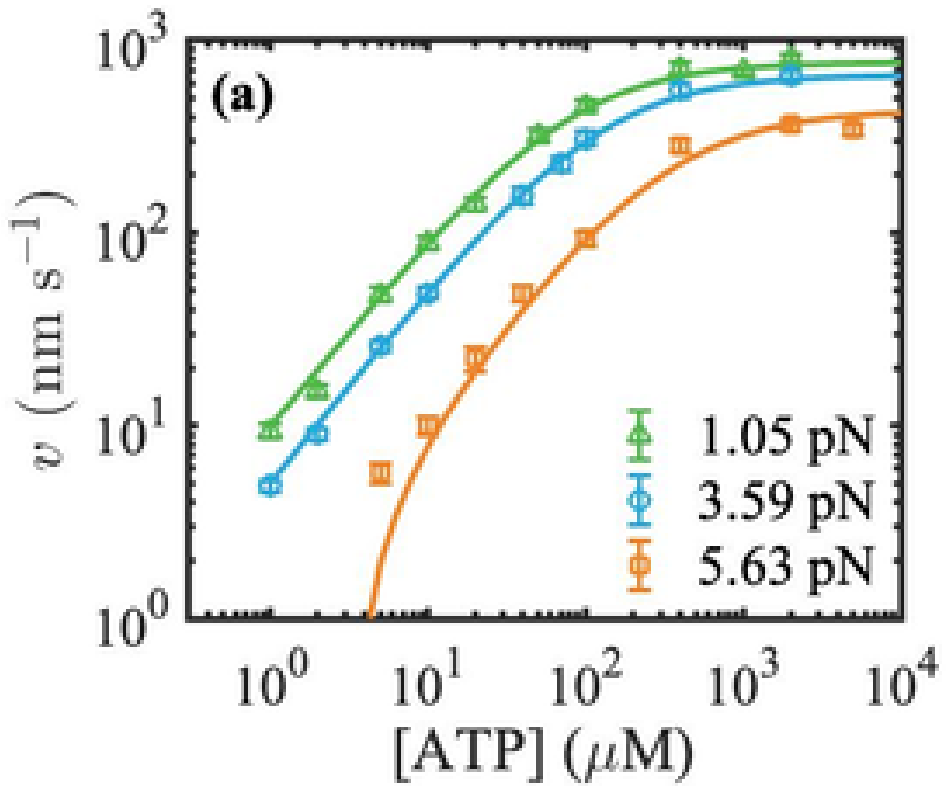}\includegraphics[scale=0.28]{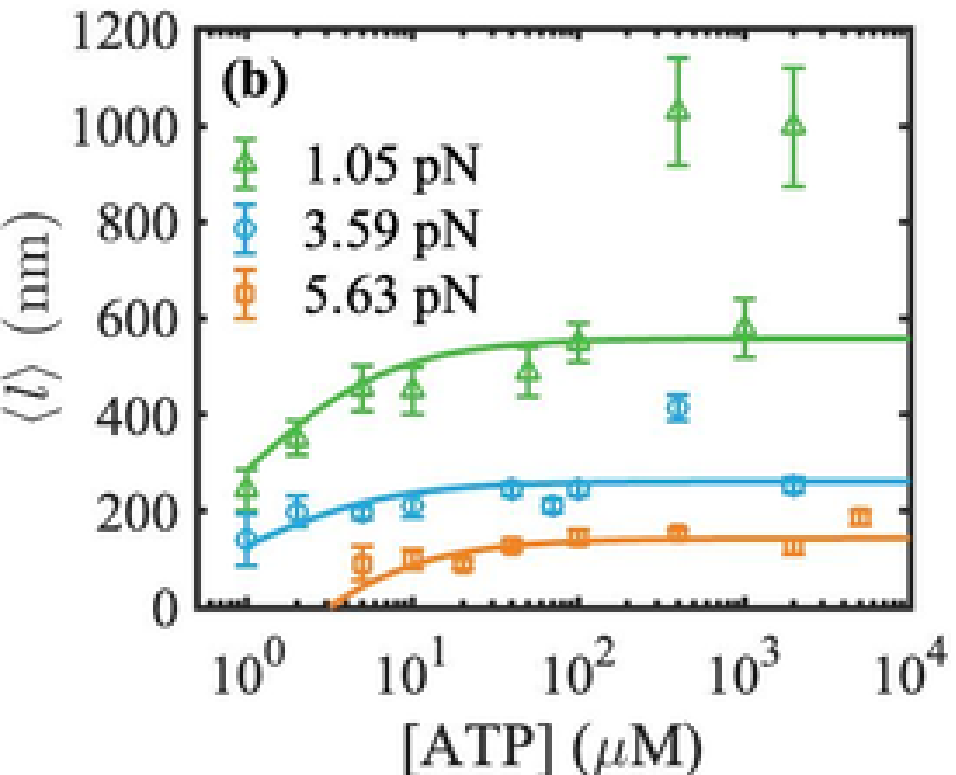}\includegraphics[scale=0.28]{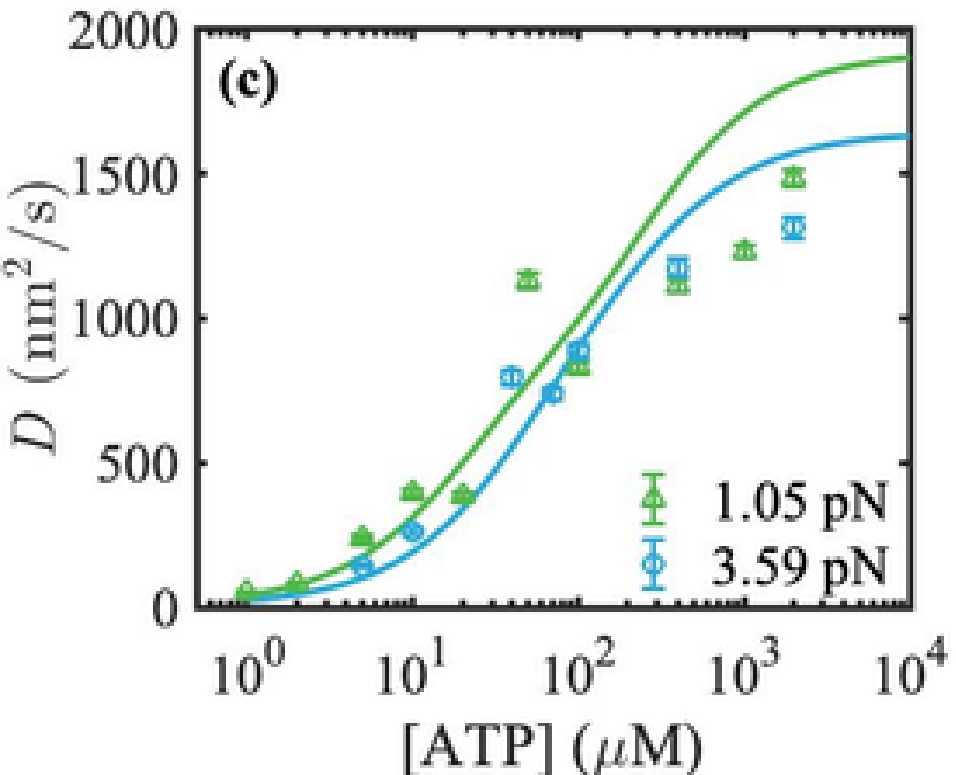}\\
\includegraphics[scale=0.28]{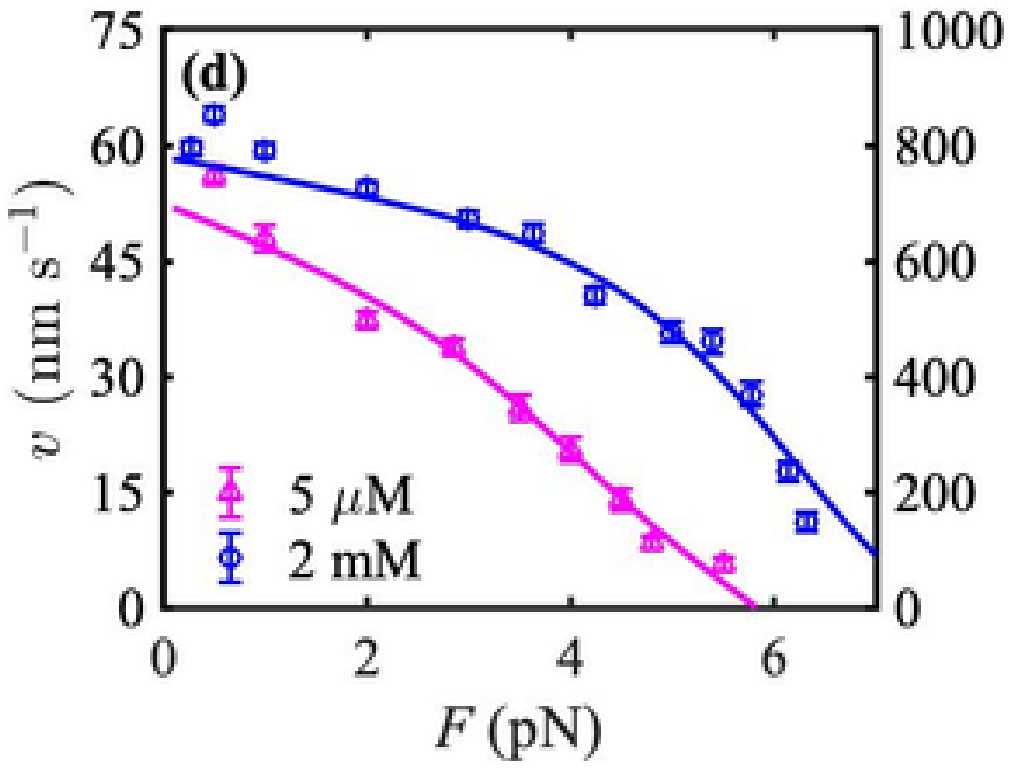}\includegraphics[scale=0.28]{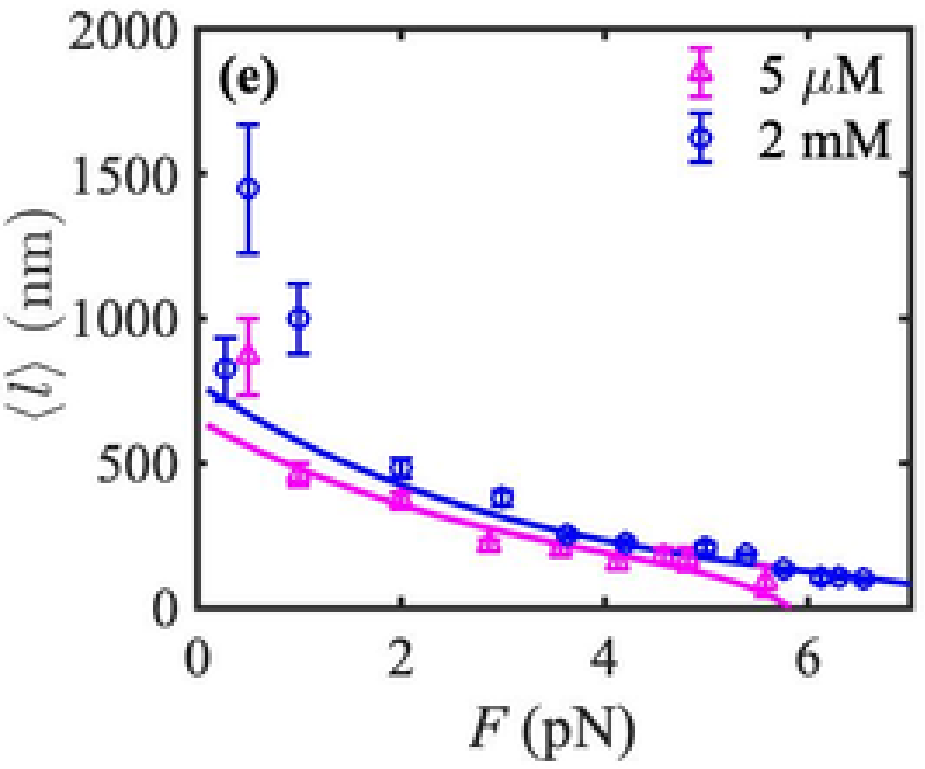}\includegraphics[scale=0.28]{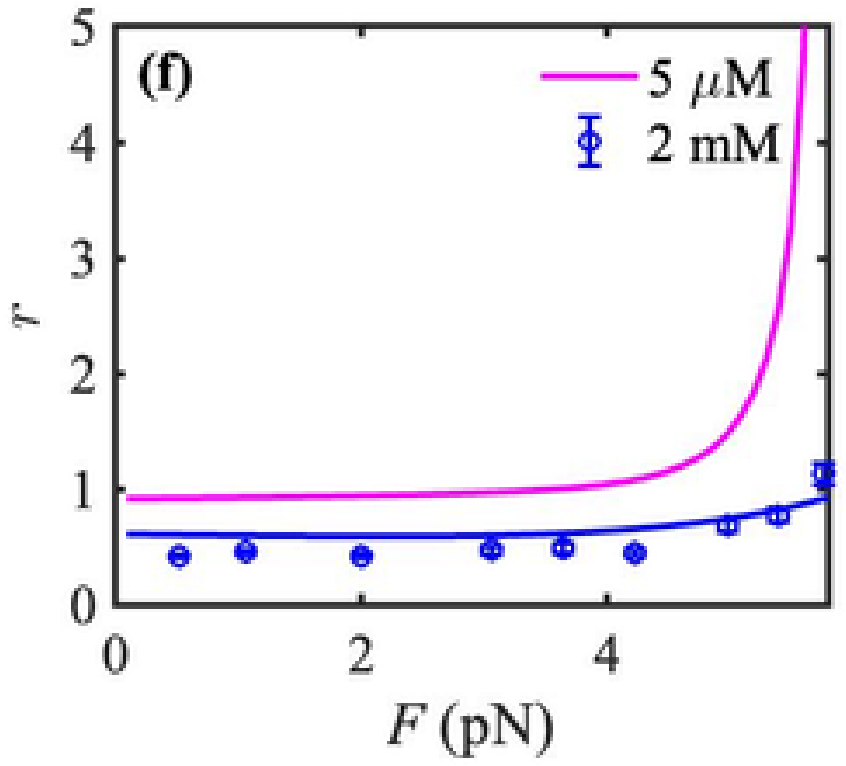}\\
	\caption{\label{fig:2} Theoretical predictions (solid lines) and experimental data (markers) of various biophysical properties of kinesin-1 purified from wild-type Drosophila. The data in \textbf{(a,b,d,e)} are from \cite{schnitzer2000force}, and in \textbf{(c,f)} are from \cite{visscher1999single}, with $D$ is estimated from $r = 2D/vd$. In \textbf{(d)} the left axis is for 5 $\mu$M ATP while the right axis is for 2 mM ATP. Theoretical results are obtained from formulations given in Eqs.~(\ref{vvv},\ref{vx},\ref{Eqr},\ref{EqRunlength}), with model parameters listed Tab.~\ref{tab:1}.}
\end{figure}

\section{Mechanochemical properties of kinesin-1}
\subsection{Biophysics of kinesin motion along MT}
Using parameter values listed in Tab.~\ref{tab:1}, our model can reproduce the related experimental results well, see Fig.~\ref{fig:2}. In the following, we will always use parameter values listed in Tab.~\ref{tab:1} to discuss the mechanochemical properties of kinesin-1.
\begin{table}[b]
\caption{\label{tab:1} Model parameter values obtained by fitting to experimental data of kinesin-1 purified from wild-type Drosophila measured in \cite{schnitzer2000force,visscher1999single}, see Figs.~\ref{fig:1} and \ref{fig:2}. }
	\begin{ruledtabular}
		\begin{tabular}{lclc}
			\textrm{Parameter}&
			\textrm{Value}&
			\textrm{Parameter}&
			\textrm{Value}\\
			\colrule
			$k_{12}^{0},k_{34}^{0}$ & 9.827 $\mu$M$^{-1}\cdot$s$^{-1}$&	$k_{21},k_{43}$ & 5074.875 s$^{-1}$\\
			$k_{23}^{0}$& 1627.099 s$^{-1}$& $\delta _{23}$& 0.055\\
			$k _{32}^{0}$&0.006 s$^{-1}$& $\delta _{32}$ & 0.416\\
			$k _{31}^{0},k _{42}^{0}$& 137.582 s$^{-1}$&$\delta _{31}$& 0\\
			$\delta _{42}$& 0.006 & $k _{13}^{0},k _{24}^{0}$& 1.666 s$^{-1}$\\
			$\delta _{13}$& 0.529 & $\delta _{24}$& 0.523 \\
			$k _{25}$& 5146.371 s$^{-1}$& $k _{52}$ &77.252 s$^{-1}$\\
			$k_{15}$&4.344 s$^{-1}$& $k_{51}$& 1455.909 s$^{-1}$\\
			$k_{56}$& 345.215 s$^{-1}$& $k_{65}$ & 2923.252 s$^{-1}$\\
			$k_{61}^0$& 0.001 s$^{-1}$& $k_{\rm off}^0$ &5.286  s$^{-1}$\\
			$d_{\rm{off}}$&0.803 nm&  &  \\
		\end{tabular}
	\end{ruledtabular}
\end{table}

From Tab.~\ref{tab:1}, $\delta_{23}+\delta_{32}\approx0.47$, which implies that there may be two substeps within each 8 nm step of kinesin, with size 3.76 nm and 4.23 nm respectively, as previously observed in \cite{Coppin1996,Nishiyama2001}. According to Fig.~\ref{fig:1}\textbf{(a)}, these two substeps are caused by  binding of the nucleotide-free head to the front site of MT and detachment of the ADP bound rear head from  MT, respectively. But, for high ATP concentration, rates $k_{12}, k_{34}$ and $k_{23}$ are very large, the substep coupled with transition $2\rightleftharpoons3$ is difficult to be observed.

The load distribution factor $\delta_{31}=0$ means the energy barrier between state 3 and 1 is very close to state 3. Transition rate $k_{31}$ is independent of external load $F$ while $k_{13}$ increases rapidly with $F$. The process $4\rightleftharpoons2$ is similar since $\delta_{42}=0.006$ is also very small.

\begin{figure}[htbp]
\includegraphics[scale=0.28]{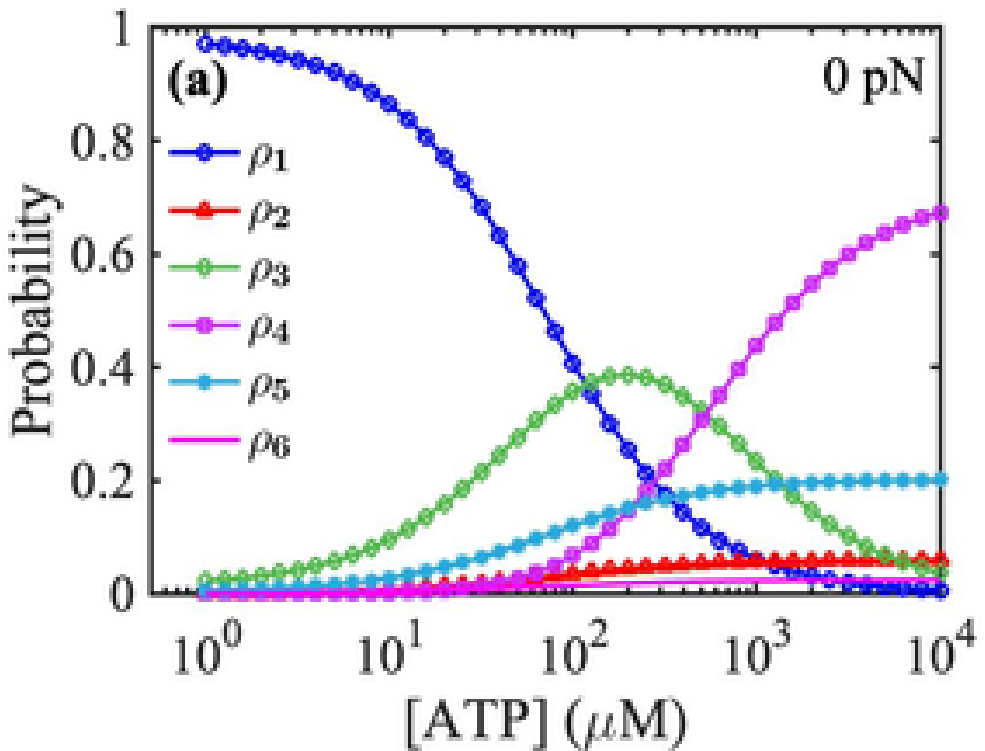}\includegraphics[scale=0.28]{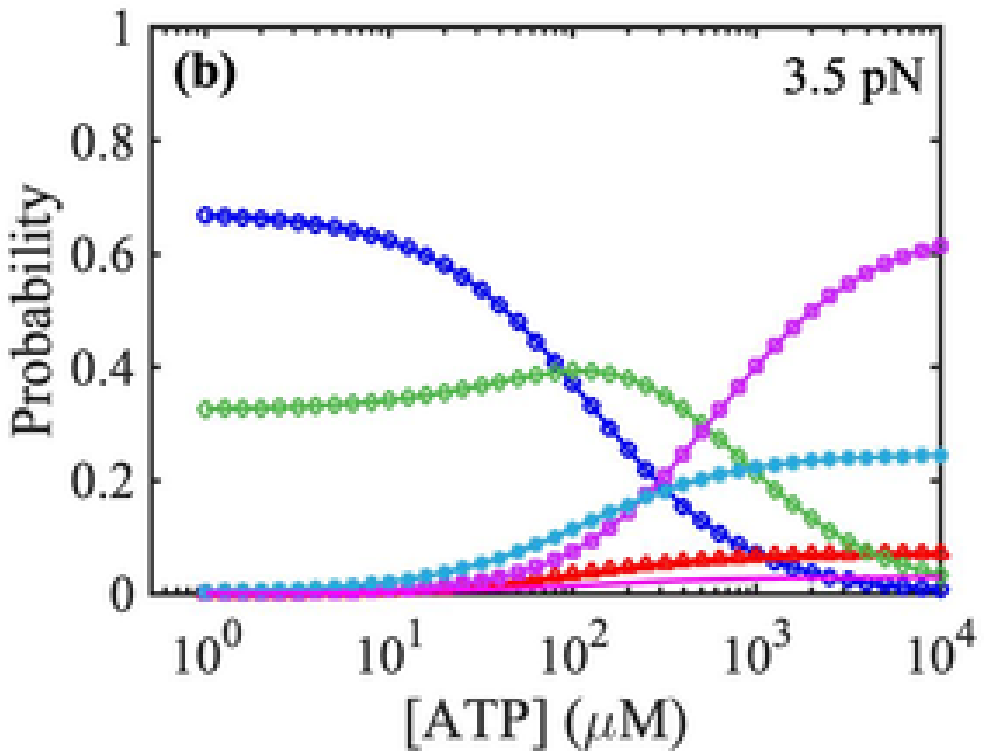}\includegraphics[scale=0.28]{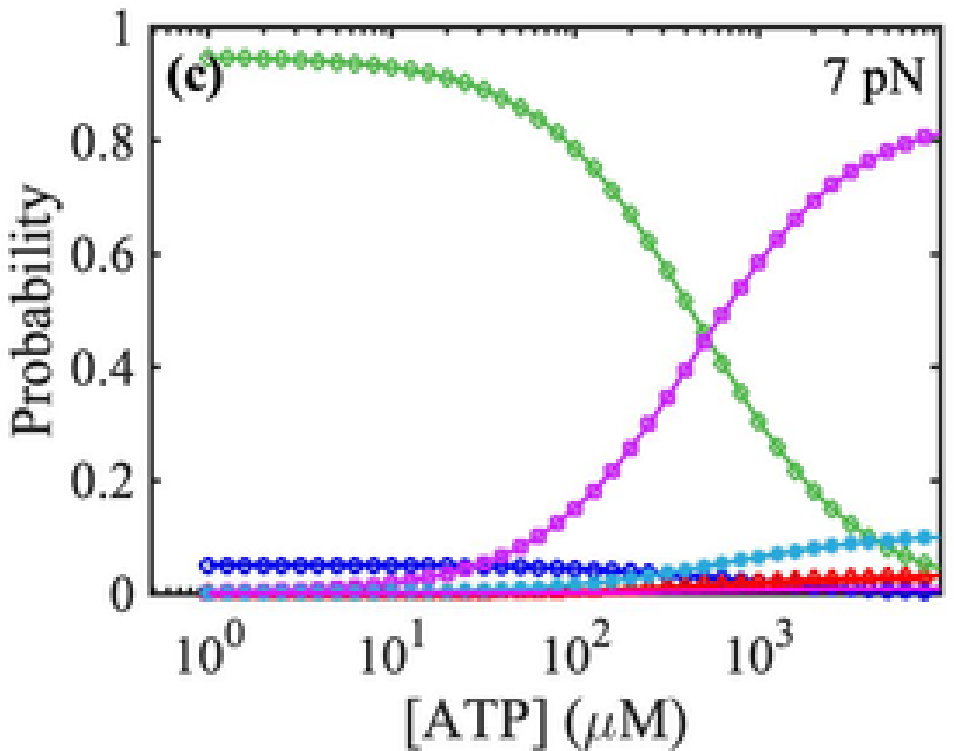}\\
\includegraphics[scale=0.28]{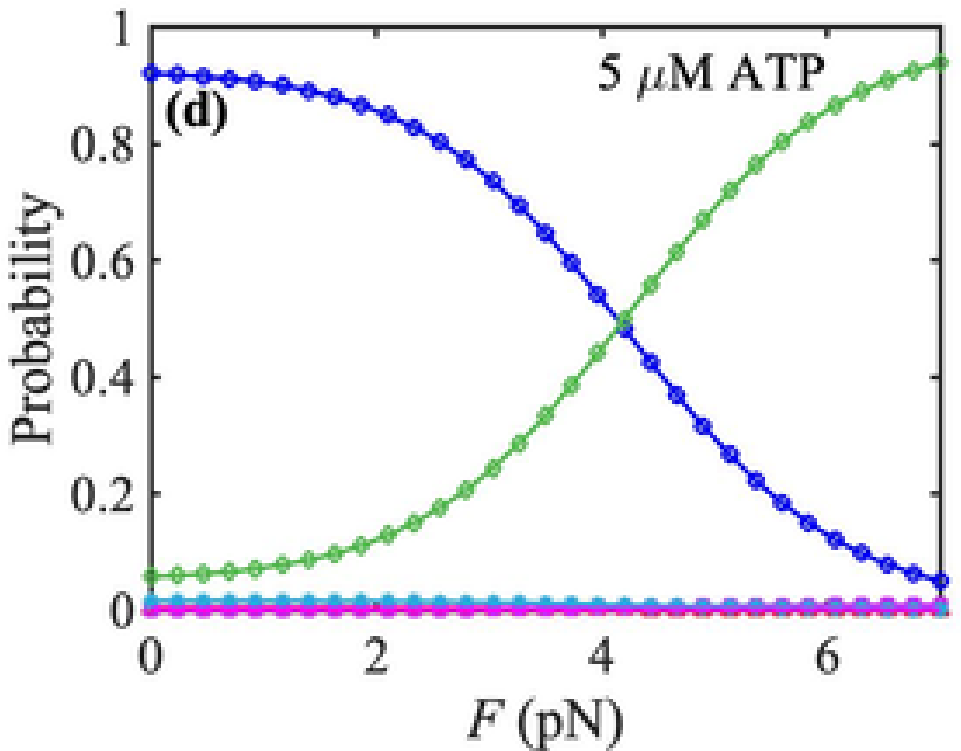}\includegraphics[scale=0.28]{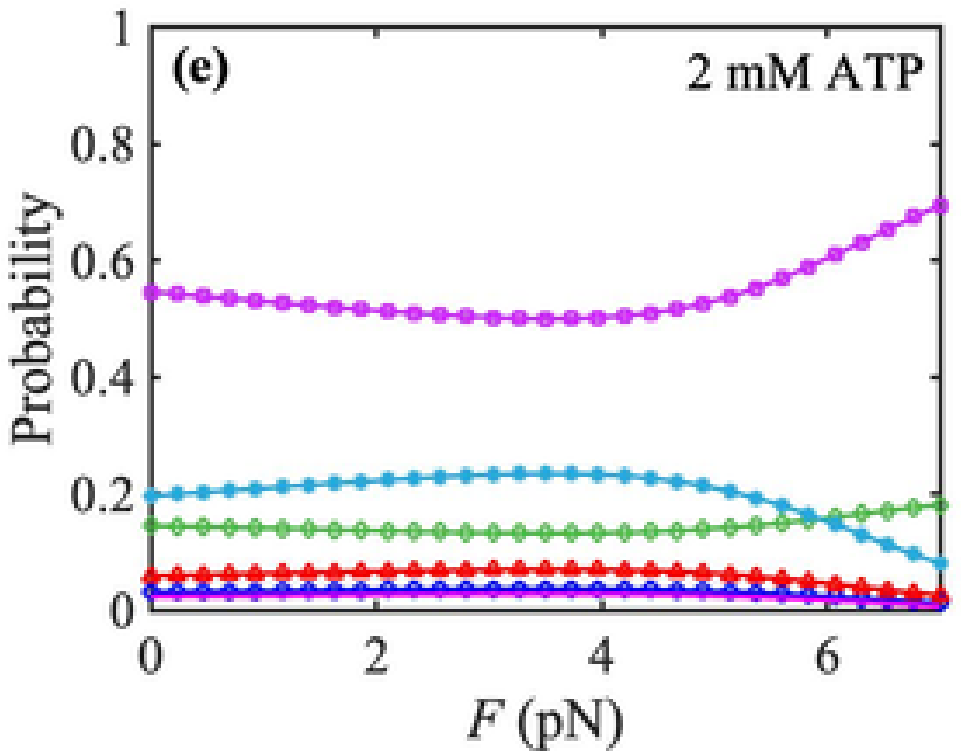}\includegraphics[scale=0.28]{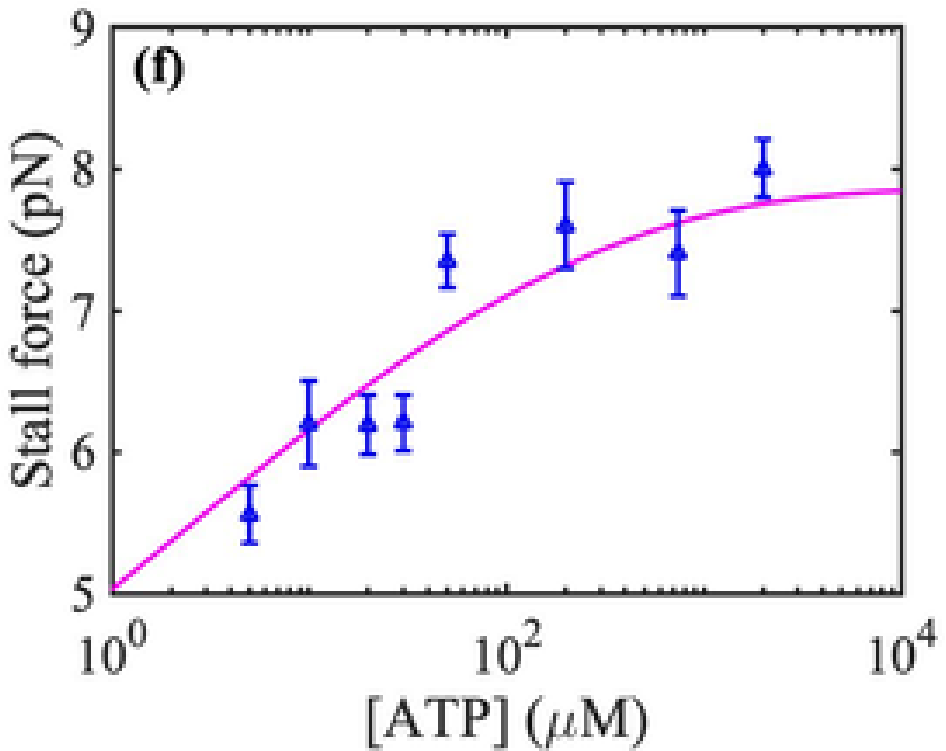}\\
\caption{\label{fig:3}\textbf{(a-c)} Probability $\rho_i$ of kinesin in the six biochemical states as depicted in Fig.~\ref{fig:1}, versus ATP concentration [ATP] at different load. \textbf{(d-e)} Probability $\rho_i$ of kinesin versus load $F$ at different ATP concentration. \textbf{(f)} Stall force of kinesin as a function of [ATP], where solid line shows the theoretical predictions using parameters listed in Tab.~\ref{tab:1}, and triangles denote experimental data measured in \cite{visscher1999single}.}
\end{figure}

Figs.~\ref{fig:3}\textbf{(a,d)} show that, at low load $F$ and low ATP concentration [ATP], kinesin mainly stays in biochemical state 1, {\it i.e.}, with only one nucleotide-free head bound to MT while the other ADP bound head detached. At high ATP concentration, kinesin stays mainly in state 4, with both heads in ATP or ADP$\cdot$Pi bound state and bound to MT tightly. Meanwhile, if the load $F$ is not high, the probability that kinesin in the two ADP bound state 5 is also relatively high (about 1/5), see Figs.~\ref{fig:3}\textbf{(a,b,e)}. At low [ATP] while high load $F$, kinesin stays mainly in state 3 with both two heads bound to MT but one in nucleotide-free state, see Figs.~\ref{fig:3}\textbf{(c,d)}. Provided the load $F$ is high, kinesin stays mainly in two head bound states, and stays mainly in state 4 when [ATP] is high while in state 3 otherwise, see Figs.~\ref{fig:3}\textbf{(c)}.

Generally, probability $\rho_1$ decreases while $\rho_2,\rho_4,\rho_5,\rho_6$ increases with ATP concentration. At high load $F$, $\rho_3$ decreases with [ATP] monotonically, otherwise $\rho_3$ may increase first and then decrease with [ATP]. At low ATP concentration, kinesin mainly stays in the ATP waiting state 1 or 3. But with the increase of load $F$ kinesin switches from the one head bound state 1 to the two heads bound state 3, see Figs.~\ref{fig:3}\textbf{(a-d)}. When ATP concentration is high, all probabilities $\rho_i$ change only slightly with the load $F$, see Fig.~\ref{fig:3}\textbf{(e)}.

Fig.~\ref{fig:3}\textbf{(f)} shows the stall force $F_s$, under which the mean velocity of kinesin vanishes, increases slightly with ATP concentration, and eventually approaching to a constant between 7 and 8 pN, which agrees well with experimental measurements \cite{Coppin1997,visscher1999single,fisher2001simple}.

\begin{figure}[htbp]
\includegraphics[scale=0.280]{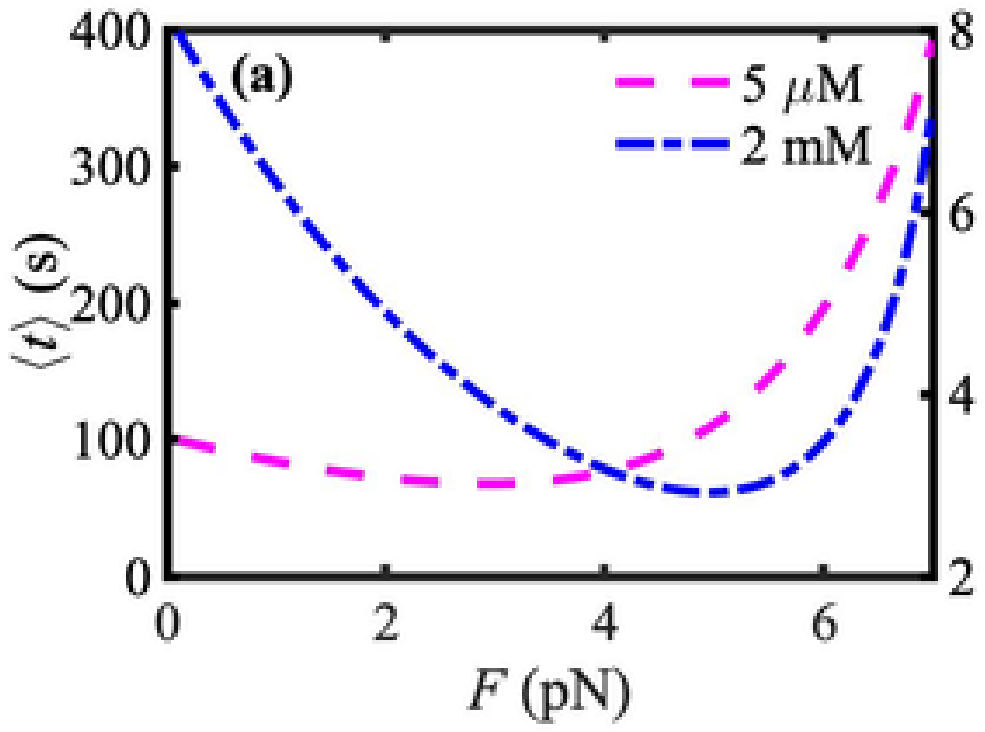}\includegraphics[scale=0.280]{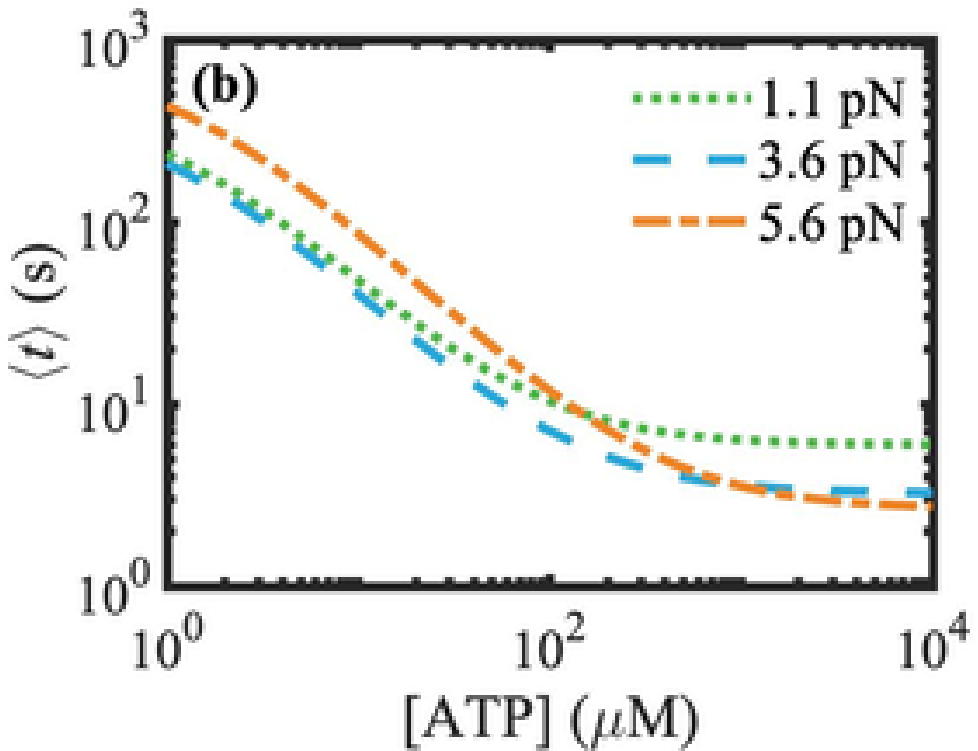}\includegraphics[scale=0.280]{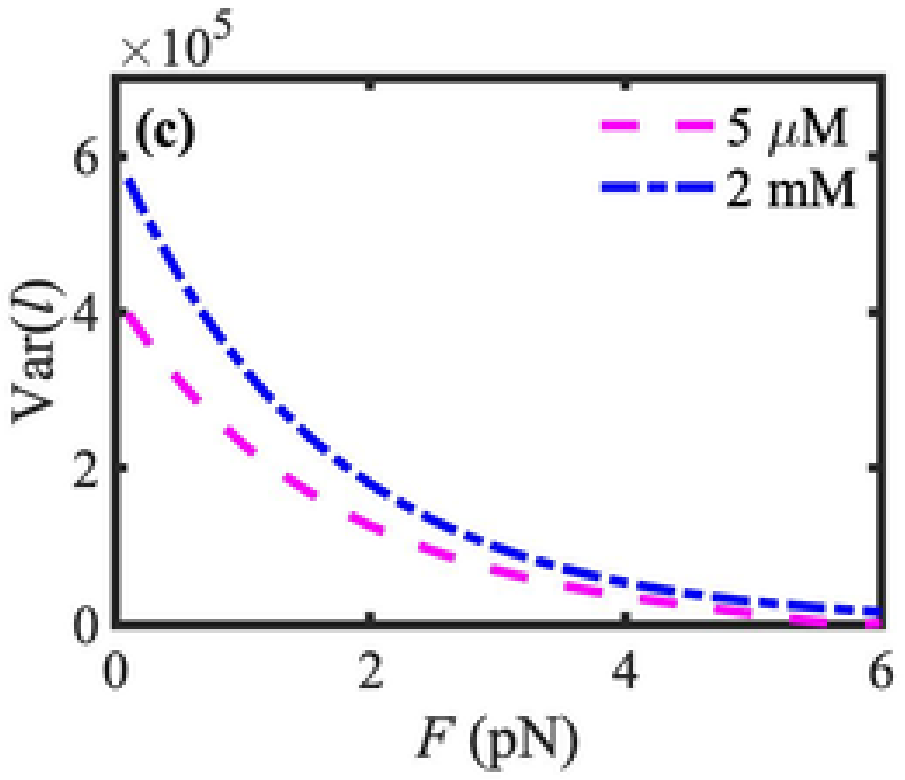}\\
\includegraphics[scale=0.280]{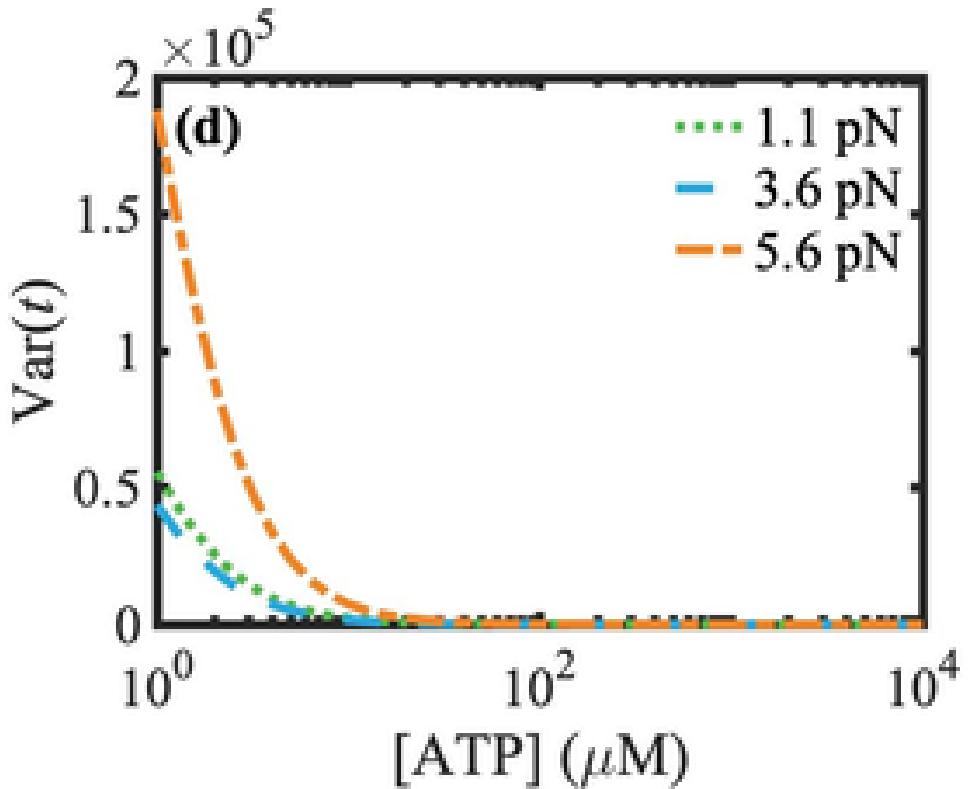}\includegraphics[scale=0.280]{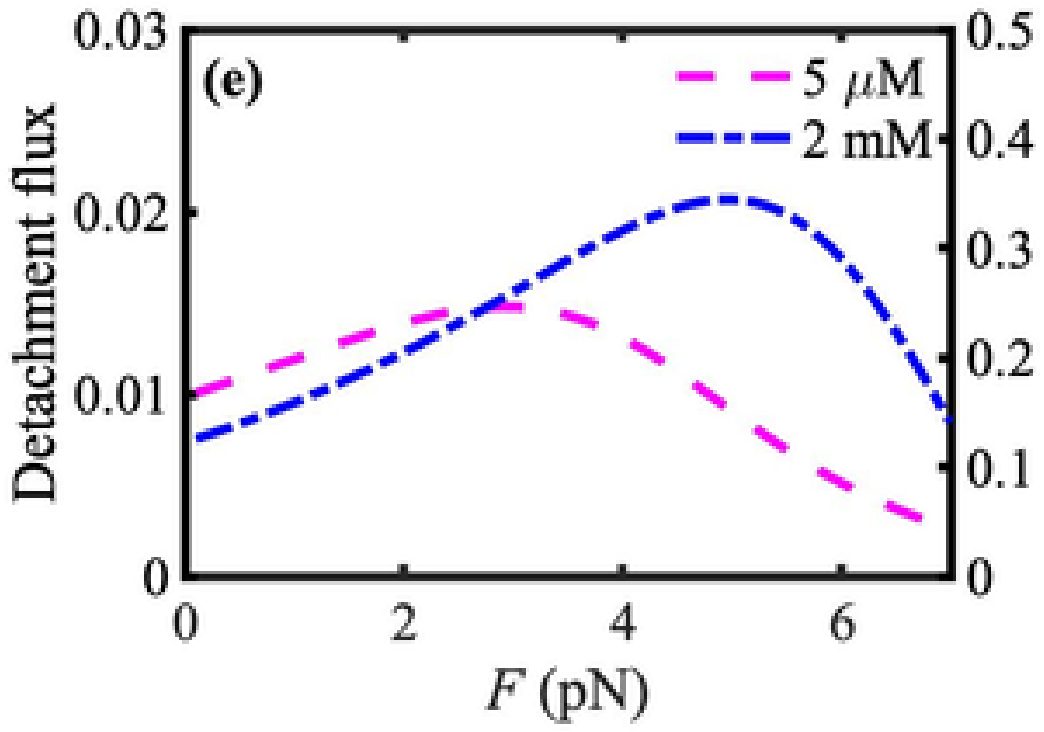}\includegraphics[scale=0.280]{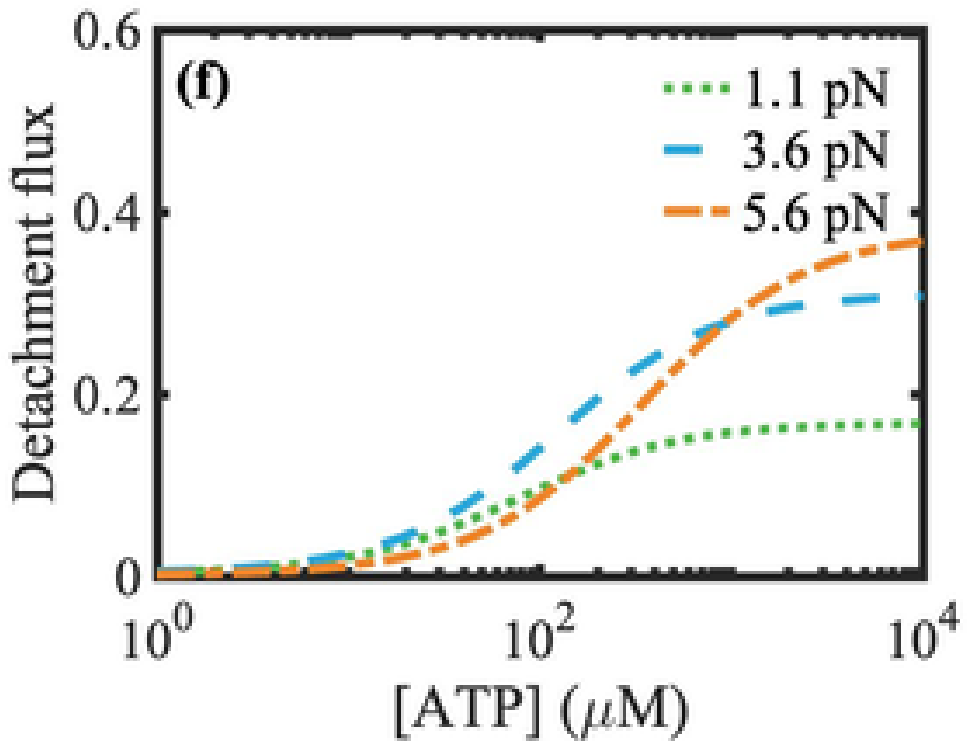}\\
\caption{\label{fig:4}\textbf{(a,b)} Mean run time $\langle t\rangle$ with the change of load $F$ and [ATP], see Eq.~(\ref{EqMeanT}). \textbf{(c)} Variance of run length ${\rm Var}(l)$ with the change of $F$, see Eq.~(\ref{EqVarRunlength}). \textbf{(d)} Variance of run time ${\rm Var}(t)$ with the change of [ATP], see Eq.~(\ref{EqVarT}). \textbf{(e,f)} Probability flux of kinesin detachment from MT, $\textrm{flux}_{\rm off}:=k_{\rm off}\rho_{6}$, with the change of load $F$ and [ATP]. In \textbf{(a,e)}, left axis is for [ATP]$=5$ $\mu$M, and right arix is for [ATP]$=2$ mM.}
\end{figure}
Figs.~\ref{fig:4}\textbf{(a,b)} show mean run time $\langle t\rangle$ of kinesin before its detachment from MT decreases with ATP concentration monotonically, while first decreases slightly and then increases rapidly with load $F$. These results seem surprising. 
One possible explanation is that, in general, kinesin does not detach from MT unless it has transported its cargo to corresponding destination \cite{Toba2006,Toprak2009}. So, with low ATP concentration, kinesin will stay on MT for more time, since the period of single biochemical cycle becomes long due to the lack of ATP molecule. Actually, Fig.~\ref{fig:4}\textbf{(b)} and Fig.~\ref{fig:2}\textbf{(b)} show both mean run time $\langle t\rangle$ and mean run length $\langle l\rangle$ tend to limit constants at high ATP concentration. One possible reason is that, in cells this run time/length is enough to transport any cargoes to their corresponding destinations. At low ATP concentration and high external load, $\langle t\rangle$  increases with $F$, which implies kinesin will try to run more time along MT to complete its task when its motion is slowed down by resistance, see Figs.~\ref{fig:4}\textbf{(a,b)}.

Generally, both mean value $\langle l\rangle$ and variance ${\rm Var}(l)$ of kinesin run length along MT decrease with load $F$, and if $F$ is not too large both $\langle l\rangle$ and ${\rm Var}(l)$ increases with ATP concentration, see Figs.~\ref{fig:2}\textbf{(b,e)}, \ref{fig:4}\textbf{(c)}, and S1\textbf{(d)}. The variance of run time ${\rm Var}(t)$ decreases with ATP concentration, while decreases first and then increases with the load $F$, see Figs.~\ref{fig:4}\textbf{(d)}, S1\textbf{(a)}. Further calculations show both coefficient of variation ${\rm CV}_l:=\langle l\rangle/\sqrt{{\rm Var}(l)}$ and ${\rm CV}_t:=\langle t\rangle/\sqrt{{\rm Var}(t)}$ almost equal to 1, see Fig.~S1\textbf{(b,c,e,f)}. So both the run time of kinesin along MT and its run length are approximately exponentially distributed.

The mean run time $\langle t\rangle$ of kinesin along MT is actually the time spent by kinesin in average to detach from MT. 
So $\langle t\rangle$ is related to the probability flux of detachment $\textrm{flux}_{\rm off}:=k_{\rm off}\rho_{6}$. Intuitively, $\textrm{flux}_{\rm off}$ should change with $F$ and [ATP] in a similar manner as $1/\langle t\rangle$. In contrast to Fig.~\ref{fig:4}\textbf{(a,b)}, the plots in Fig.~\ref{fig:4}\textbf{(e,f)} show detachment flux $\textrm{flux}_{\rm off}$ increases monotonically with [ATP], while increases first and then decreases with load $F$.

\begin{figure}[htbp]
\includegraphics[scale=0.280]{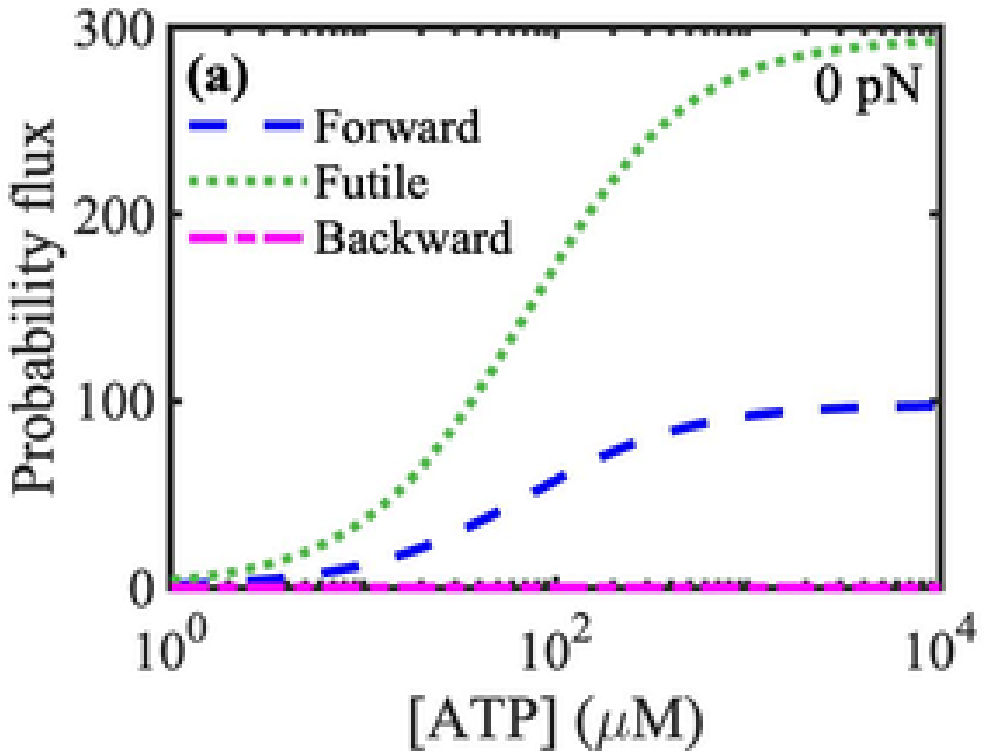}\includegraphics[scale=0.280]{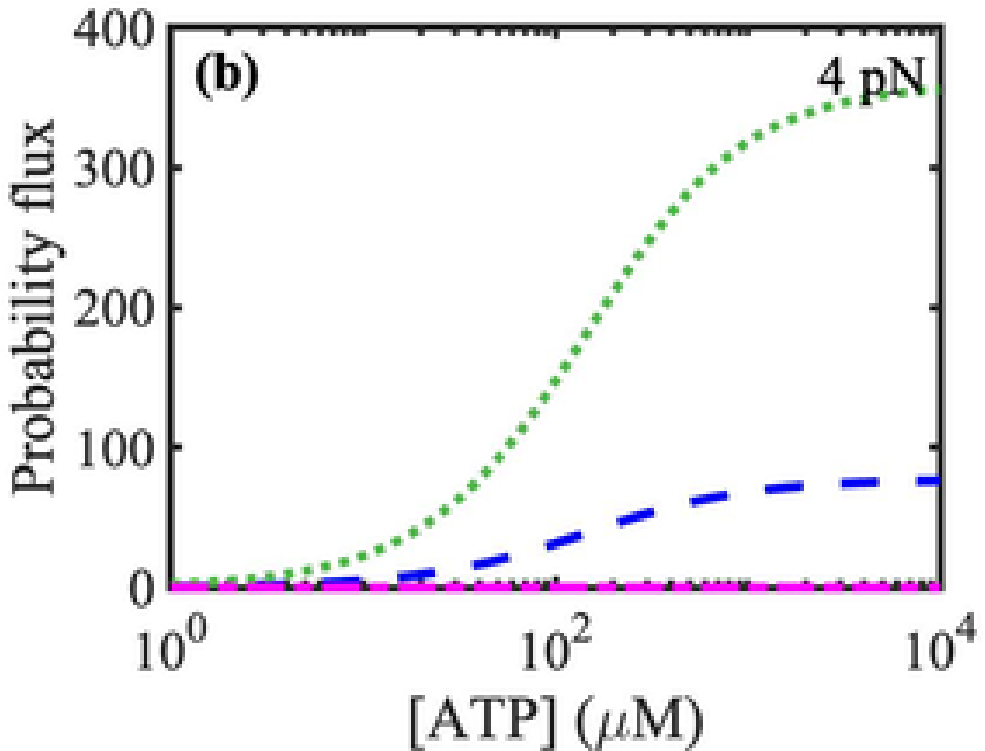}\includegraphics[scale=0.280]{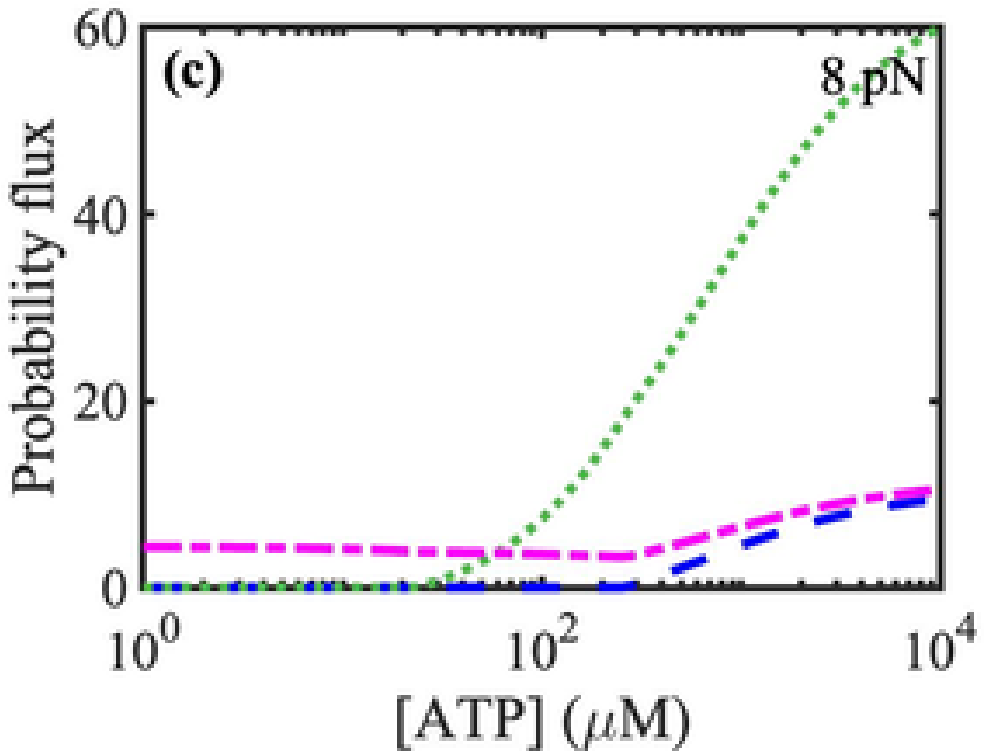}\\
\includegraphics[scale=0.280]{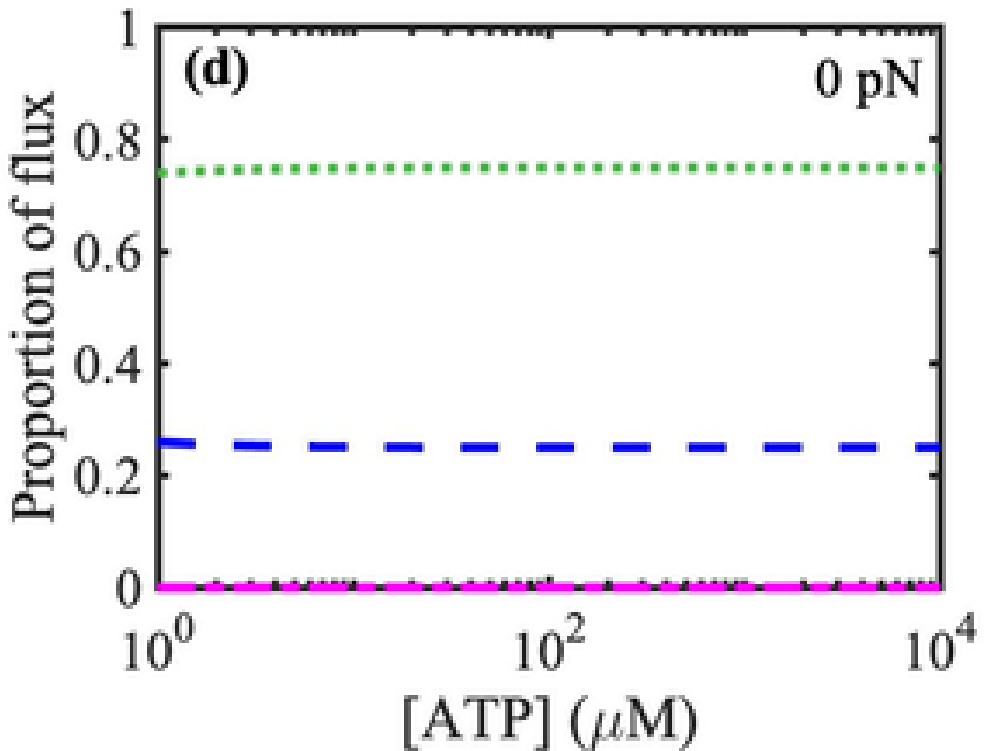}\includegraphics[scale=0.280]{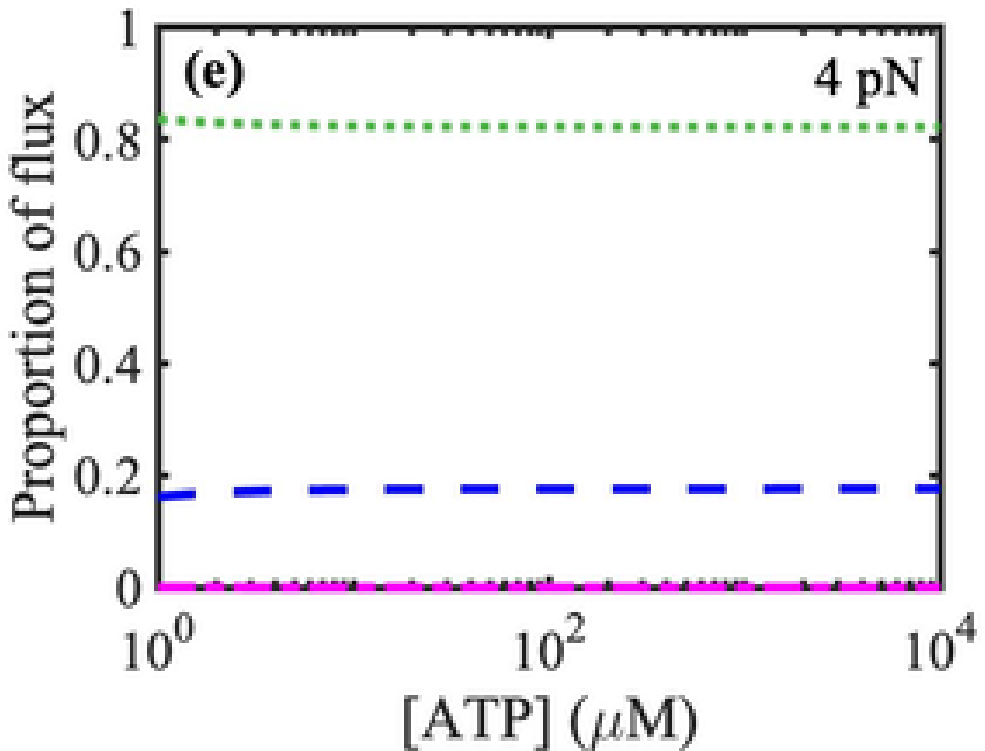}\includegraphics[scale=0.280]{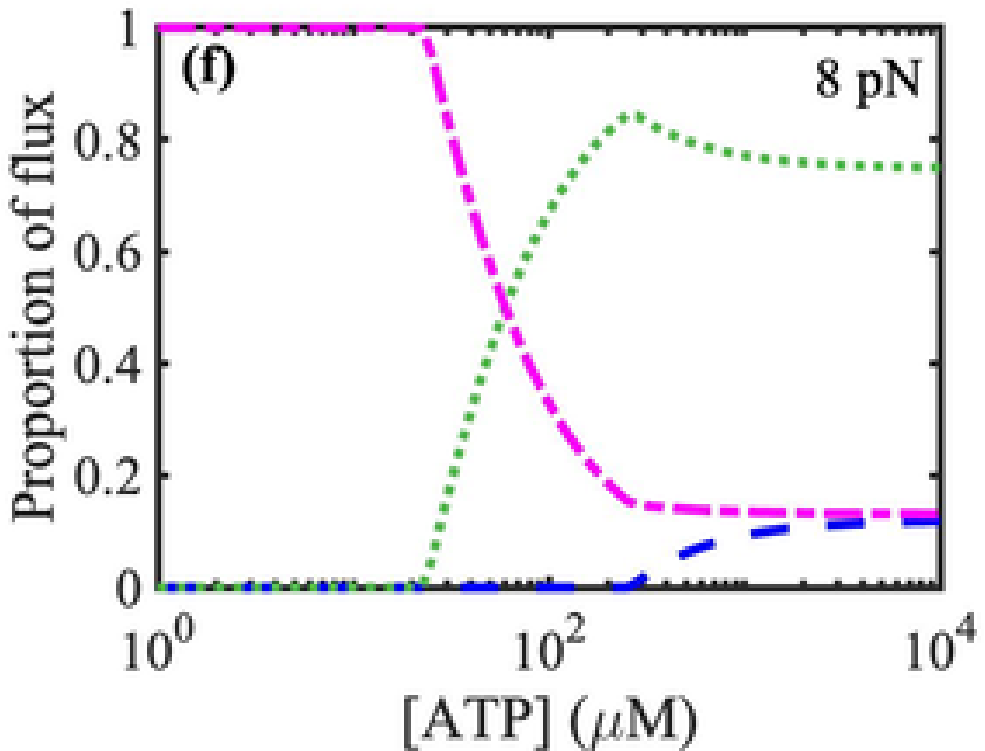}\\
\caption{\label{fig:5} Probability flux of {\it forward/backward/futile} cycle of kinesin-1, as well as their proportions, versus ATP concentration [ATP], with load $F=0$ pN, 4 pN and 8 pN, respectively. }
\end{figure}

\subsection{Pathways of forward/backward/futile biochemical cycle of kinesin-1}
Given the model depicted in Fig.~\ref{fig:1}, there is no evident information about the coupling between biochemical cycle of ATP hydrolysis and mechanical step of kinesin. As pointed out before, there are generally three categories of the biochemical cycle, {\it forward/backward/futile} cycle. Mean velocity $v$ is the sum of probability fluxes of different biochemical cycles, weighted by mechanical step. It is generally difficult to sort out these three categories of biochemical cycle and then calculate their corresponding fluxes, since net flux between any two adjacent biochemical states $i$ and $j$, $\textrm{flux}_{ij}:=k_{ij}\rho_{i}-k_{ji}\rho_{j}$, usually changes with external load $F$ and ATP concentration [ATP]. Fortunately, thorough numerical calculations show that, with parameter values listed in Tab.~\ref{tab:1} and biophysically meaningful load $F$ and [ATP], flux variation in our model for kinesin-1 consists of only 5 different cases, in each case the direction of net flux between any two adjacent states does not change, see Figs.~S2-S4. For each case, {\it forward/backward/futile} cycle can be phenomenologically sorted out by analyzing all possible biochemical pathways. The results are summarized in Tab.~\ref{tab:2}.
\begin{table}
  \centering\footnotesize
\begin{tabular}{l|l|c|c|c|c}
\hline
 Conditions & Cycles  & Pathway & ATP & Steps & Total flux \\
\hline\hline
\multirow{4}{*}{\tabincell{c}{$++++$\\ {\bf(Case 1)}}} & \multirow{2}{*}{Forward} & $\overrightarrow{1231}$ & 1 & 1 & \multirow{2}{*}{$\textrm{flux}_{23}$} \\
\cline{3-5}
     &  & $\overrightarrow{2342}$ & 1 & 1 &  \\
\cline{2-6}
     & Backward  & $\overrightarrow{12561}$ & 1 & -1 & $\textrm{flux}_{61}$ \\
\cline{2-6}
     & Futile & $\overrightarrow{1251}$  & 1  &  0 &  $\textrm{flux}_{51}$ \\
\hline\hline
\multirow{5}{*}{\tabincell{c}{$-+++$\\ {\bf(Case 2)}}} & Forward & $\overrightarrow{2342}$ & 1 & 1 & $\textrm{flux}_{23}$ \\
\cline{2-6}
     & \multirow{2}{*}{Backward}  & $\overrightarrow{12561}$ & 1 & -1 & \multirow{2}{*}{$\textrm{flux}_{61}$} \\
\cline{3-5}
     &  & $\overrightarrow{1342561}$ & 1 & -1 &  \\
\cline{2-6}
     & \multirow{2}{*}{Futile} & $\overrightarrow{1251}$  & 1  &  0 &  \multirow{2}{*}{$\textrm{flux}_{51}$} \\
\cline{3-5}
     &  & $\overrightarrow{134251}$ & 1 & 0 & \\
\hline\hline
\multirow{7}{*}{\tabincell{c}{$--++$\\ {\bf(Case 3)}}} & Forward & --- & --- & --- & $0$ \\
\cline{2-6}
     & \multirow{4}{*}{Backward}  & $\overrightarrow{12561}$ & 1 & -1 & \multirow{4}{*}{$\textrm{flux}_{61}+\textrm{flux}_{51}\frac{\textrm{flux}_{32}}{\textrm{flux}_{25}}$} \\
\cline{3-5}
     &  & $\overrightarrow{1342561}$ & 1 & -1 &  \\
\cline{3-5}
     &  & $\overrightarrow{13251}$ & 0 & -1 &  \\
\cline{3-5}
     &  & $\overrightarrow{132561}$ & 0 & -2 &  \\
\cline{2-6}
     & \multirow{2}{*}{Futile} & $\overrightarrow{1251}$ & 1  &  0 &  \multirow{2}{*}{$\textrm{flux}_{51}\frac{\textrm{flux}_{12}+\textrm{flux}_{42}}{\textrm{flux}_{25}}$} \\
\cline{3-5}
     &  & $\overrightarrow{134251}$ & 1 & 0 & \\
\hline\hline
\multirow{6}{*}{\tabincell{c}{$---+$\\ {\bf(Case 4)}}} & Forward & --- & --- & --- & $0$ \\
\cline{2-6}
     & \multirow{4}{*}{Backward}  & $\overrightarrow{12561}$ & 1 & -1 & \multirow{4}{*}{\tabincell{c}{$\textrm{flux}_{61}+\textrm{flux}_{24}$\\ +$\textrm{flux}_{51}\frac{\textrm{flux}_{13}}{\textrm{flux}_{25}}$}} \\
\cline{3-5}
     &  & $\overrightarrow{132561}$ & 0 & -2 &  \\
\cline{3-5}
     &  & $\overrightarrow{13251}$ & 0 & -1 &  \\
\cline{3-5}
     &  & $\overrightarrow{2432}$ & -1 & -1 &  \\
\cline{2-6}
     & Futile & $\overrightarrow{1251}$ & 1  &  0 &  $\textrm{flux}_{51}\frac{\textrm{flux}_{12}}{\textrm{flux}_{25}}$ \\
\hline\hline
\multirow{6}{*}{\tabincell{c}{$----$\\ {\bf(Case 5)}}} & Forward & --- & --- & --- & $0$ \\
\cline{2-6}
     & \multirow{4}{*}{Backward}  & $\overrightarrow{1321}$ & -1 & -1 & \multirow{4}{*}{$\textrm{flux}_{32}+\textrm{flux}_{61}$} \\
\cline{3-5}
     &  & $\overrightarrow{132561}$ & 0 & -2 &  \\
\cline{3-5}
     &  & $\overrightarrow{2432}$ & -1 & -1 &  \\
\cline{3-5}
     &  & $\overrightarrow{13251}$ & 0 & -1 &  \\
\cline{2-6}
     & Futile & --- & ---  &  --- &  $0$ \\
\hline
\end{tabular}
\caption{Pathway details of {\it forward/backward/futile} biochemical cycle of kinesin-1. Column \lq Conditions' lists the signs of $\textrm{flux}_{31}, \textrm{flux}_{23}, \textrm{flux}_{34}$, and $\textrm{flux}_{12}$, respectively. Column \lq ATP' lists the number of ATP molecule consumed in the corresponding pathway, where \lq$-1$' means one ATP is synthesized. Column \lq Steps' gives the number of mechanical step coupled with the pathway, with one step 8 nm. Column \lq Total flux' lists the total flux of the {\it forward/backward/futile} cycle. $\protect\overrightarrow{ij\cdots ki}$ denotes the biochemical pathway $i\!\to\! j\!\to\!\cdots\!\to\! k\!\to\! i$. It is evident from Fig.~\ref{fig:1} that $\textrm{flux}_{42}=\textrm{flux}_{34}$. Flux $\textrm{flux}_{25}$, $\textrm{flux}_{56}=\textrm{flux}_{61}$, and $\textrm{flux}_{51}$ are always nonnegative for $0\le F\le 9$ pN and 1 $\mu$M $\le$ [ATP] $\le$ 10000 $\mu$M, see Fig.~S2. }\label{tab:2}
\end{table}

Results show that, at low load $F$ the flux of {\it backward} cycle $\textrm{flux}_-$ is negligible, indicating the impossibility of backward motion \cite{Toba2006,Toprak2009}, see Figs.~\ref{fig:5}\textbf{(a,b,d,e)} and Fig.~S5. Both {\it forward} flux $\textrm{flux}_+$ and {\it futile} flux $\textrm{flux}_0$ increase with ATP concentration [ATP]. At low load $F$, proportions of $\textrm{flux}_+$ and $\textrm{flux}_0$ are almost independent of [ATP], which implies that the utilization ratio of ATP is independent of [ATP] at low load, see Figs.~\ref{fig:5}\textbf{(d,e)}. At super stall load, kinesin motion is dominated by {\it backward} cycle when [ATP] is small, while dominated by {\it futile} cycle when [ATP] is large. Particularly, for $F=8$ pN, the proportion of {\it backward} flux $\textrm{flux}_-$ is almost 1 when [ATP] $<22.23$ $\mu$M, indicating that kinesin-1 moves backward almost deterministically.  For large enough [ATP], the lower limit value of the proportion of $\textrm{flux}_-$ is almost the same as the upper limit value of the proportion of $\textrm{flux}_+$, so the motion of kinesin is completed stalled, see Figs.~\ref{fig:5}\textbf{(c,f)}. With increasing [ATP], the proportion of $\textrm{flux}_0$ rises rapidly, until 263.47 $\mu$M, at which the sign of $\textrm{flux}_{23}$ changes from negative to positive,  switching from {\bf Case 3} to {\bf Case 2}, producing the flux of {\it forward} cycle $\textrm{flux}_+>0$, see Fig.~S3\textbf{(c)} and Figs.~\ref{fig:5}\textbf{(c,f)}. On the whole, at high load and low ATP concentration, kinesin-1 can only move backward. As [ATP] increases, the futile flux begins to rise rapidly, and then forward flux is generated. For large enough [ATP], {\it backward} flux is balanced by {\it forward} flux, and the motion of kinesin-1 is stalled completely.

Further calculations show that both {\it forward} flux $\textrm{flux}_+$ and its proportion decrease, while the proportion of $\textrm{flux}_-$ increases, with load $F$, which implies the utilization ratio of ATP decreases with load $F$, see Fig.~S5. At [ATP] = 5 $\mu$M and [ATP] = 2 mM, the intersection points of $\textrm{flux}_+$ and $\textrm{flux}_-$ are found at $F=5.85$ pN and $F=7.76$ pN, respectively, which are exactly the stall forces $F_s$ of kinesin at [ATP] = 5 $\mu$M and [ATP] = 2 mM, see Fig.~\ref{fig:3}(f).
One surprising result is that, except the extreme cases with high load $F$ and very low [ATP], the biochemical process of kinesin-1 is mainly dominated by {\it futile} cycle, and utilization ratio of ATP is only about 20\%, see Fig.~\ref{fig:5}. At small load $F$, the {\it backward} flux is almost zero. While for load $F$ larger than stall force, the {\it forward} flux decreases rapidly to zero. This indicates that, unless at external load which is near the stall force, the motion of kinesin-1 is almost deterministic, and at small load, kinesin-1 hardly steps backward. This is different with the conclusion draw from Brownian ratchet models \cite{Astumian1997,Zhang2008}.

Tab.~\ref{tab:2} shows that {\it forward} cycle consists of two possible pathways,  $\overrightarrow{1231}$ and  $\overrightarrow{2342}$, both of them include hydrolysis of one ATP and a forward mechanical step of 8 nm. {\it Futile} cycle consists of two possible pathways,  $\overrightarrow{1251}$ and  $\overrightarrow{134251}$, both of them include hydrolysis of one ATP but without mechanical step. {\it Backward} cycle consists of six possible pathways, which are $\overrightarrow{12561}$, $\overrightarrow{1342561}$,  $\overrightarrow{13251}$, $\overrightarrow{132561}$, $\overrightarrow{2432}$ and $\overrightarrow{1321}$.
These six pathways can be further classified into three categories, {\bf (i)} $\overrightarrow{12561}$ and $\overrightarrow{1342561}$, which are mainly induced by backward sliding 6$\to$1 through the semi-detach state, {\bf (ii)} $\overrightarrow{13251}$, $\overrightarrow{2432}$, and $\overrightarrow{1321}$, which can be roughly regarded as the reversal of corresponding {\it forward} pathways, and {\bf (iii)} $\overrightarrow{132561}$, which includes both the reversal of a forward pathway and the backward sliding 6$\to$1, and therefore coupled with a backward motion of total 16 nm. Pathway $\overrightarrow{2432}$ and $\overrightarrow{1321}$ include the synthesis of ATP, while $\overrightarrow{12561}$ includes the hydrolysis of ATP. Note that pathway $\overrightarrow{ij\cdots ki}$ can be denoted equivalently as $\overrightarrow{j\cdots kij}$, $\overrightarrow{kij\cdots k}$, etc. For convenience, we denote the probability flux through pathway $\overrightarrow{ij\cdots ki}$ by $f_{\overrightarrow{ij\cdots ki}}$.

With small [ATP] and load $F\ll F_s$, the {\it forward} cycle is mainly realized through pathway $\overrightarrow{1231}$, see Figs.~S6\textbf{(a,d)}, S8\textbf{(a,d)}, and S9\textbf{(a,d)}. Otherwise, the {\it forward} cycle is mainly realized through pathway $\overrightarrow{2342}$, Figs.~S6-S10\textbf{(a,d)}. During pathway $\overrightarrow{2342}$, binding of ATP to the front head of kinesin is earlier than the release of phosphate Pi from the rear head, while during pathway $\overrightarrow{1231}$, ATP binds to the front head only after the detachment of rear head from MT. So our results imply that at low external load and low ATP concentration, the MT bound head might be in the nucleotide-free state, while at high ATP concentration or high load, the head bound to MT is always in ATP or ADP$\cdot$Pi binding state. This is consistent with results shown in Fig.~\ref{fig:3}.

Except some extreme cases, the {\it futile} cycle is mainly realized through the pathway $\overrightarrow{1251}$. Moreover, for small load $F$ or small [ATP], the probability that the {\it futile} cycle realized through $\overrightarrow{134251}$ is almost negligible, see Figs.~S6-S10\textbf{(b,e)}.

If load $F$ is small, the {\it backward} cycle is mainly realized through pathway $\overrightarrow{12561}$, and other pathways are all negligible, see Figs.~S6-S9\textbf{(c,f)}. If both $F$ and [ATP] are large, the {\it backward} cycle is mainly realized through pathway $\overrightarrow{12561}$ and pathway $\overrightarrow{1342561}$, and also with $\overrightarrow{12561}$ the most prominent one, see Figs.~S7\textbf{(c,f)} and S10\textbf{(c,f)}. If $F$ is large but [ATP] is small, pathways $\overrightarrow{13251}$, $\overrightarrow{2432}$, $\overrightarrow{1321}$ and $\overrightarrow{132561}$ are all important, while both pathway $\overrightarrow{12561}$ and pathway $\overrightarrow{1342561}$ are negligible, see Figs.~S6\textbf{(c,f)} and S10\textbf{(e,f)}. For large $F$ and small [ATP], prominent pathways of {\it backward} cycle are roughly related to the reversal of corresponding pathways of {\it forward} cycle, and consequently might lead to the synthesis of ATP, see Tab.~\ref{tab:2}. However, this is difficult to observe experimentally, since kinesin-1 may have already detached from MT before achieving such harsh conditions.
\begin{figure}[htbp]
\includegraphics[scale=0.280]{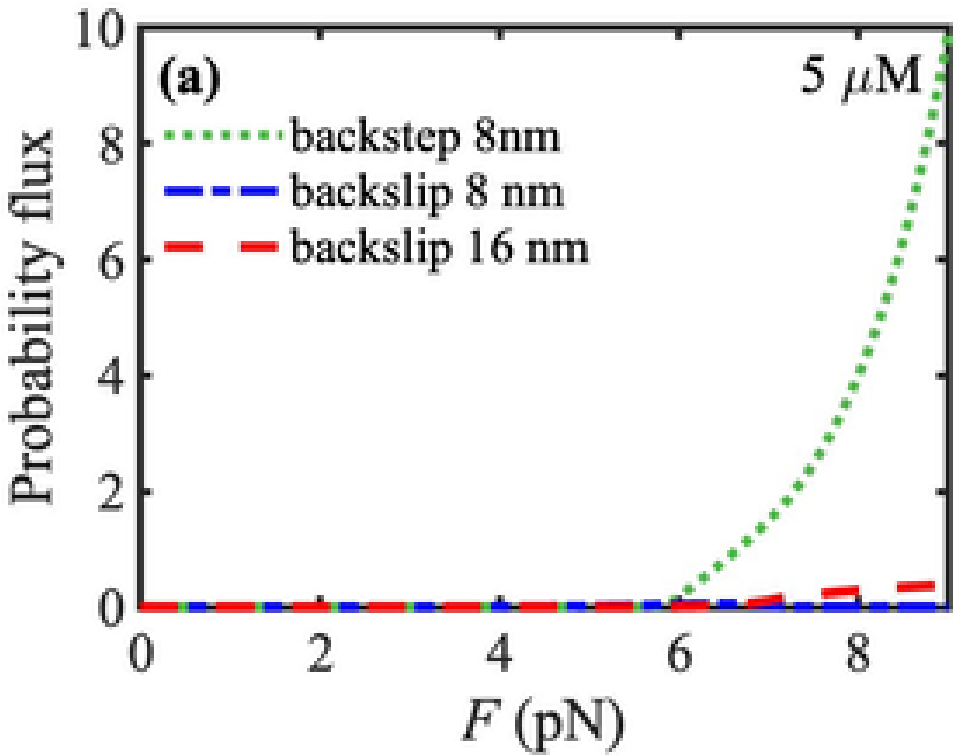}\includegraphics[scale=0.280]{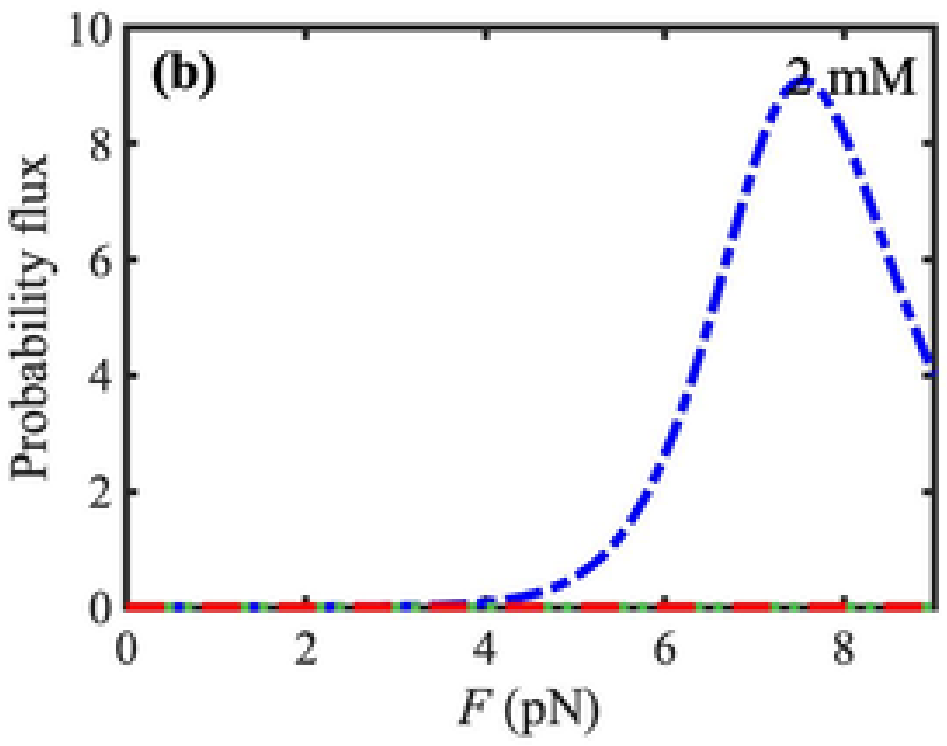}\includegraphics[scale=0.280]{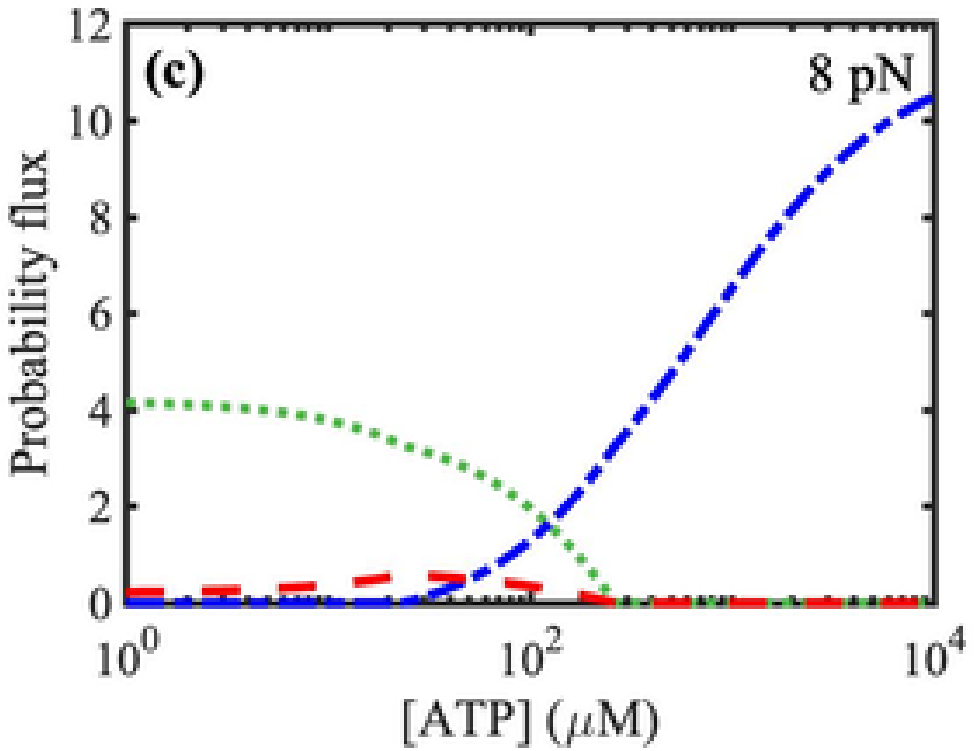}\\
\includegraphics[scale=0.280]{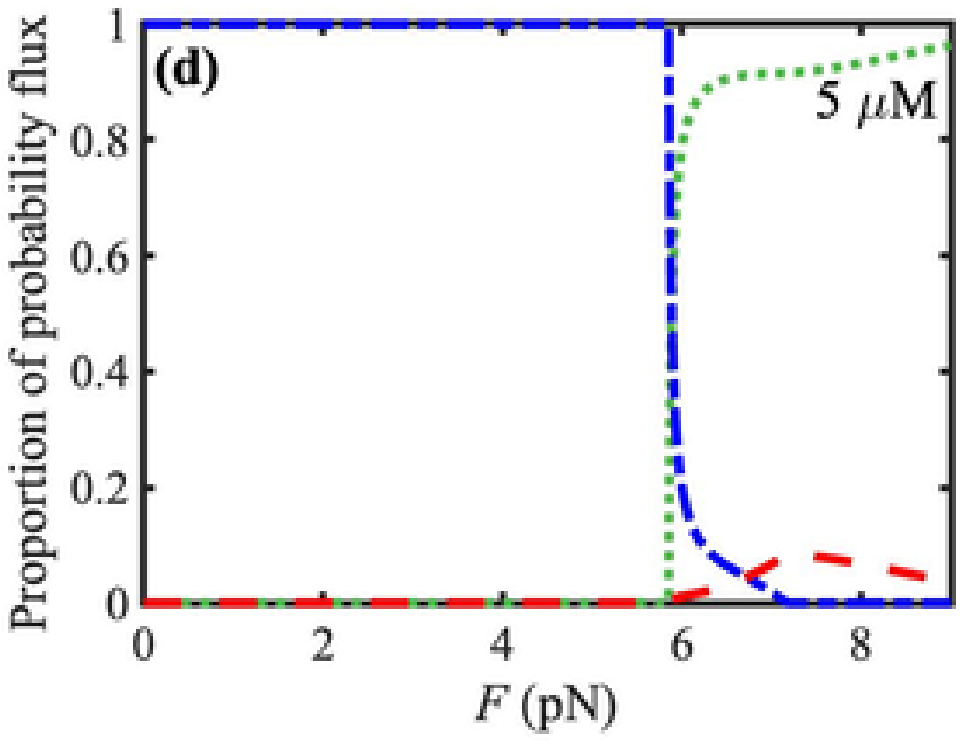}\includegraphics[scale=0.280]{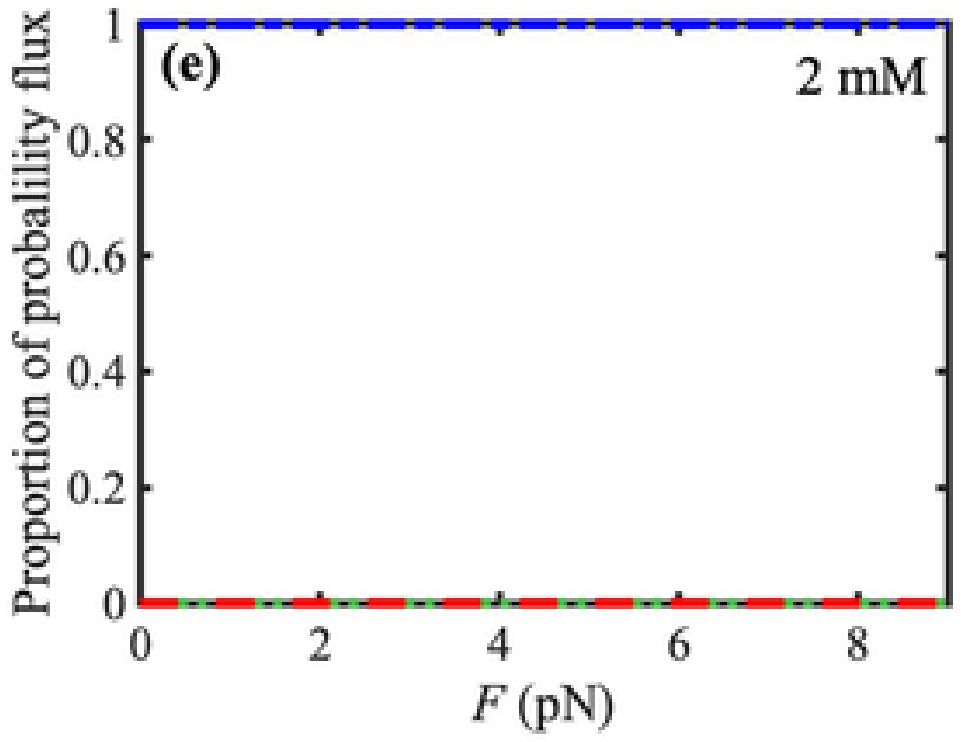}\includegraphics[scale=0.280]{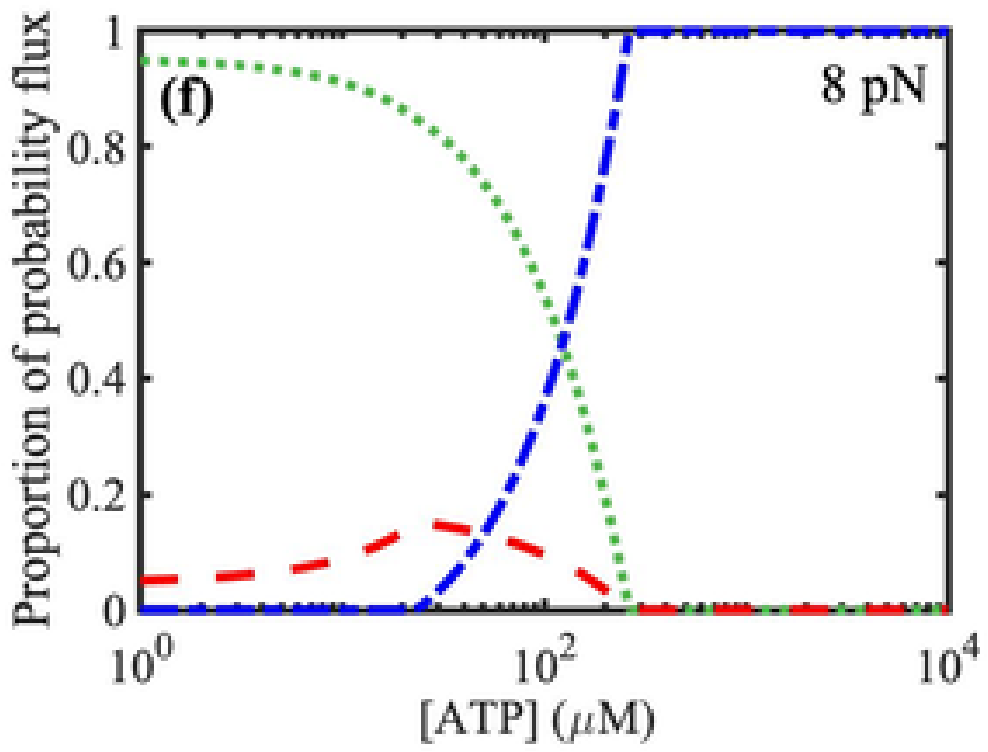}\\
\caption{\label{fig:6} Probability flux of 8 nm backstep, 8 nm backslip and 16 nm backslip, as well as their corresponding proportions at various [ATP] and load $F$. \textbf{(a-c)} show the probability flux at [ATP] = 5 $\mu$M, [ATP] = 2 mM, and $F=8$ pN, respectively, and \textbf{(d-f)} show their corresponding proportions. The flux of 8 nm backstep is obtained by $f_{\protect\overrightarrow{13251}}+f_{\protect\overrightarrow{2432}}+f_{\protect\overrightarrow{1321}}$, the flux of 8 nm backslip is obtained by $f_{\protect\overrightarrow{12561}}+f_{\protect\overrightarrow{1342561}}$, and the flux of 16 nm backslip is $f_{\protect\overrightarrow{132561}}$.
}
\end{figure}
During pathways $\overrightarrow{12561}$ and $\overrightarrow{1342561}$, backward motion of kinesin-1 is induced only by the backward sliding of 8 nm through transition 6$\to$1. But during pathway $\overrightarrow{132561}$, the backward motion is accomplished by both one backward sliding of 8 nm and one backward step of 8 nm. Fig.~\ref{fig:6} shows that the probability flux through pathway $\overrightarrow{132561}$ is very rare and the backward step of 16 nm is observable only under the very harsh condition that external load $F$ is high but ATP is scarce. If ATP concentration reach saturating, both probability flux of 8 nm backstep and 16 nm backslip decrease to zero, only backslip 8nm existing, see Figs.~\ref{fig:6}{\bf (b,c)}.\\

\section{Conclusions and remarks}
According to recent experiments, a mechanochemical model of kinesin-1 is constructed, in which the motor can step forward/backward through multiple biochemical pathways. Biophysical quantities of kinesin, including its mean velocity, diffusion constant, and mean run time/length along microtubule, can be obtained theoretically.

Our study shows that mean run time of kinesin before its detachment from microtubule decreases with ATP concentration monotonically and finally tends to an external load-dependent limit value, but decreases first and then increases rapidly with external load, which are different with properties of mean run length observed in experiments. However, both distributions of run time and run length are approximately exponential.

Forward motion of kinesin-1 can be realized through two biochemical pathways. At high ATP concentration or under high external load, the kinesin head bound to microtubule will always be in ATP or ADP$\cdot$Pi binding state. But at low ATP concentration and low load, new ATP molecule will not bind to the nucleotide-free head before the release of phosphate Pi and the detachment of the trailing head from microtubule. Kinesin-1 may stay at the one head bound state for a long time to wait for ATP arrival.

Backward motion of kinesin-1 is mainly caused by backward sliding along microtubule through semi-detach state, and hardly through directional reversal of forward pathways as usually believed in previous studies. Large backward step and directional reversal of forward pathways can only happen in very harsh environments with scarce ATP molecules but under high external load. Usually, kinesin-1 has already been pulled down from microtubule before achieving such conditions.

One surprising finding is that the utilization ratio of ATP is only about 20\%, and about 80\% of ATP is consumed in futile cycle, which consists of two possible biochemical pathways. Unless ATP concentration is extremely low but external load is high, ATP is required in almost all backward steps of kinesin-1, though these backward steps are mostly accomplished through the sliding process in its semi-detach state.


\end{document}